\documentclass[aps,prc,superscriptaddress,twocolumn,showpacs,nofootinbib]{revtex4-1}
\usepackage[utf8]{inputenc}
\usepackage[T1]{fontenc}
\usepackage{graphicx}   
\usepackage{latexsym}   
\usepackage{enumerate}
\usepackage{rotating,booktabs,multirow}
\usepackage{amsmath}
\usepackage{amsfonts}
\usepackage{amssymb}
\usepackage{multirow}
\usepackage{bm}
\usepackage{subfigure}
\usepackage{xcolor}
\usepackage[colorlinks=true,linktocpage=true,linkcolor=blue,citecolor=blue,allcolors=blue]{hyperref}

\newcommand{\tmo}{t_\mathrm{overlap}}

\begin{document}
\title{Untangling the interplay of the Equation-of-State and the Collision Term towards the generation of Directed and Elliptic Flow at intermediate energies}

\author{Tom Reichert}
\affiliation{Institut f\"ur Theoretische Physik, Goethe-Universit\"at Frankfurt, Max-von-Laue-Strasse 1, D-60438 Frankfurt am Main, Germany}
\affiliation{Frankfurt Institute for Advanced Studies (FIAS), Ruth-Moufang-Str.1, D-60438 Frankfurt am Main, Germany}
\affiliation{Helmholtz Research Academy Hesse for FAIR (HFHF), GSI Helmholtz Center for Heavy Ion Physics, Campus Frankfurt, Max-von-Laue-Str. 12, 60438 Frankfurt, Germany}

\author{J\"org~Aichelin}
\affiliation{SUBATECH UMR 6457 (IMT Atlantique, Universit\'e de Nantes, IN2P3/CNRS), 4 Rue Alfred Kastler, F-44307 Nantes, France}
\affiliation{Frankfurt Institute for Advanced Studies (FIAS), Ruth-Moufang-Str.1, D-60438 Frankfurt am Main, Germany}

\begin{abstract}
The mechanism for generating  directed and elliptic flow in heavy-ion collisions is investigated and quantified for the SIS18 and SIS100 energy regimes. The observed negative elliptic flow $v_2$, at midrapidity has been explained either via (in-plane) shadowing or via (out-of-plane) squeeze-out. To settle this question, we employ the Ultra-relativistic Quantum Molecular Dynamics model (UrQMD) to calculate Au+Au collisions at E$_\mathrm{lab}=0.6A$ GeV, E$_\mathrm{lab}=1.23A$ GeV and $\sqrt{s_\mathrm{NN}}=3.0$ GeV using a hard Skyrme type Equation-of-State to calculate the time evolution and generation of directed flow and elliptic flow. We quantitatively distinguish the impact of collisions and of the potential on $v_1$ and $v_2$ during the evolution of the system. These calculations reveal that in this energy regime the generation of $v_1$ and $v_2$ follows from a highly intricate interplay of different processes and is created late, after the system has reached its highest density and has created a matter bridge between projectile and target remnant, which later breaks. Initially, we find a strong out-of-plane pressure. Then follows a strong stopping and the built up of an in-plane pressure. The $v_2$, created by both processes, compensate to a large extend. The finally observed $v_2$ is caused by the potential, reflects the freeze-out geometry and can neither be associated to squeeze-out nor to shadowing. 

  
The results are highly relevant for experiments at GSI, RHIC-FXT and the upcoming FAIR facility, but also for experiments at FRIB, and strengthens understanding on the Equation-of-State at large baryon densities.
\end{abstract}

\maketitle

\section{Introduction}
Strongly interacting nuclear matter, theoretically described by Quantum Chromo Dynamics (QCD), can experimentally be investigated in collisions of protons or heavy-ions in the worlds largest particle accelerators like the LHC at CERN, RHIC at BNL or SIS18/100 at GSI/FAIR. The experiments carried out at these accelerator facilities yield insights into the structure of QCD \cite{NA49:1999myq,STAR:2002eio,PHENIX:2003nhg,Gazdzicki:2008kk,ALICE:2008ngc,HADES:2009aat}. 

Strongly interacting matter can also be studied by analyzing observations of astrophysical objects, such as neutron stars or their binary mergers \cite{Most:2022wgo,Jakobus:2023fru}. The observation of very heavy neutron stars with more than twice the solar mass \cite{Miller:2019cac,Riley:2019yda,Miller:2021qha,Riley:2021pdl} lead to the conclusion that neutron star cores must be very stiff. The Equation-of-State (EoS) of highly compressed matter is the driving quantity determining the relationship between the maximal radius and the maximal mass of a neutron star. Thus a precise knowledge of the EoS is required to understand very heavy neutron stars \cite{Ozel:2010bz,Bonanno:2011ch,Lastowiecki:2011hh,Blaschke:2015uva}. Two and three body hyperon+nucleon interactions may play a role there \cite{Haidenbauer:2019boi,Haidenbauer:2021wld}, or more exotic phases \cite{Blaschke:2021poc,Shahrbaf:2022upc} may be produced. 

With the advent of gravitational wave detection \cite{LIGOScientific:2018cki,LIGOScientific:2020aai,LIGOScientific:2020zkf} a third source of information on the EoS may appear but these studies are still at the very beginning.

The study of neutron stars by satellite \cite{Miller:2019cac,Riley:2019yda,Miller:2021qha,Riley:2021pdl}, further observations of gravitational waves from neutron star collisions  \cite{Bauswein:2012ya,Most:2018eaw,Most:2022wgo,Jakobus:2023fru} as well as improved detector systems at the upcoming new FAIR facility near Darmstadt, will enable physicists to provide further, more precise data in the Equation-of-State, both for symmetric and asymmetric matter. 

\begin{figure} [t!hb]
    \centering
    \includegraphics[width=\columnwidth]{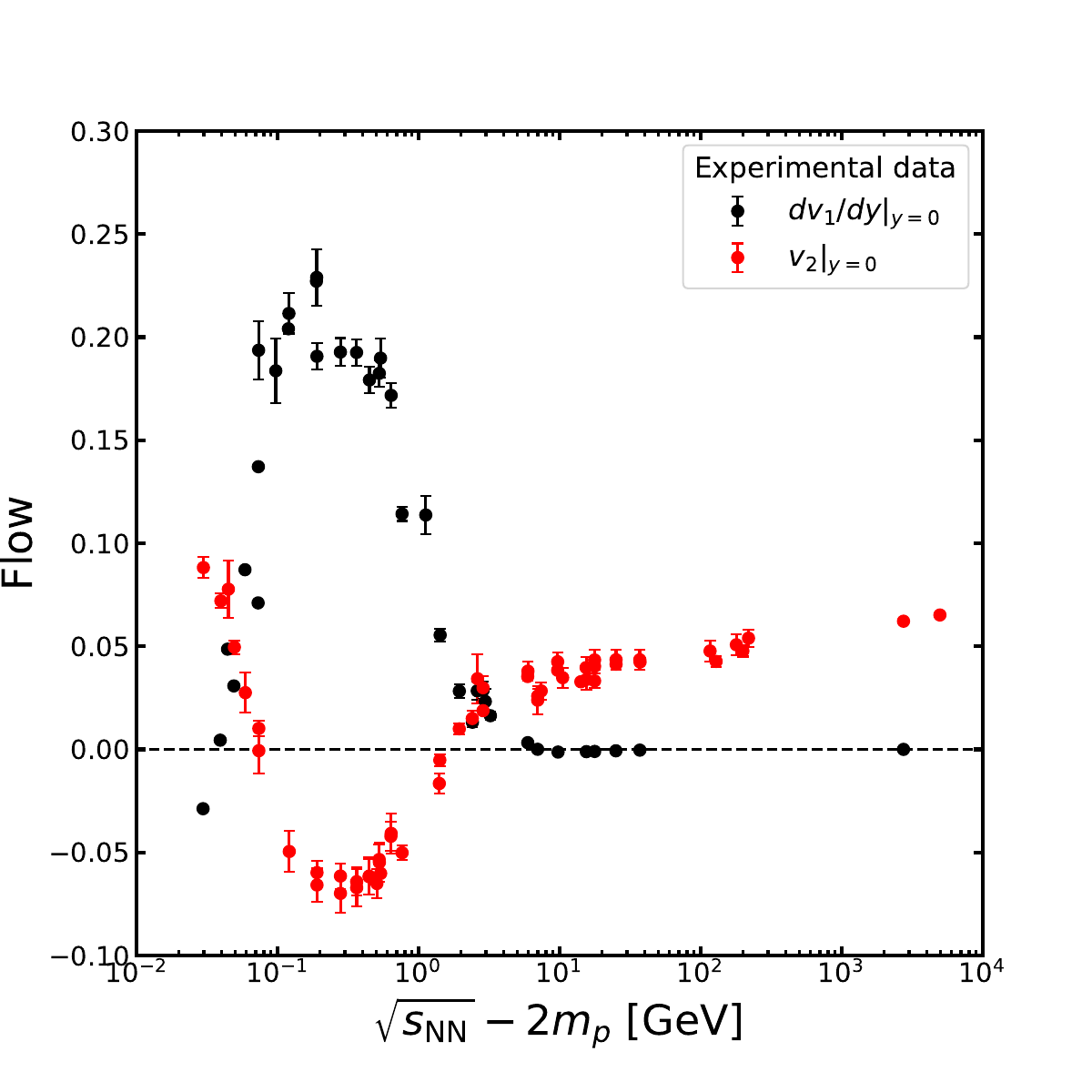}
    \caption{[Color online] Summary of experimental data on the midrapidity slope of directed flow $\mathrm{d}\langle v_1\rangle /\mathrm{d}y|_{y=0}$ and the elliptic flow at midrapidity $\langle v_2\rangle |_{y=0}$ as a function of the reduced center of mass energy $\sqrt{s_\mathrm{NN}}-2m_p$. The figure has been adapted from \cite{HADES:2022osk} and the original experimental data are from Refs. \cite{Andronic:2006ra,Andronic:2001sw,Andronic:2004cp,Andronic:2006ra,Andronic:2004cp,FOPI:2011aa,Pinkenburg:1999ya,Liu:2000am,E877:1997zjw,Alt:2003ab,Adamczyk:2014ipa,STAR:2020dav,STAR:2021ozh,Kashirin:2020evw,Barrette:1994xr,E877:1996czs,Adamova:2002qx,Aggarwal:2004zh,Abelev:2009bw,Adamczyk:2012ku,Back:2004zg,Doss:1987kq,Gutbrod:1989wd,HADES:2022osk}.}
    \label{fig:experimental_flow}
\end{figure}

In heavy-ion collisions the Equation-of-State is usually studied  by the elliptic flow $v_2$ and, complementary, by the directed flow $v_1$ \cite{Voloshin:1994mz,Sorensen:2023zkk,Sahu:1999mq}. In Fig. \ref{fig:experimental_flow} we show the summarized experimental data on the midrapidity slope of directed flow $\mathrm{d}\langle v_1\rangle /\mathrm{d}y|_{y=0}$ and the elliptic flow at midrapidity $\langle v_2\rangle |_{y=0}$ as a function of the reduced center of mass energy $\sqrt{s_\mathrm{NN}}-2m_p$. The figure has been adapted from \cite{HADES:2022osk} and the original experimental data is from Refs. \cite{Andronic:2006ra,Andronic:2001sw,Andronic:2004cp,Andronic:2006ra,Andronic:2004cp,FOPI:2011aa,Pinkenburg:1999ya,Liu:2000am,E877:1997zjw,Alt:2003ab,Adamczyk:2014ipa,STAR:2020dav,STAR:2021ozh,Kashirin:2020evw,Barrette:1994xr,E877:1996czs,Adamova:2002qx,Aggarwal:2004zh,Abelev:2009bw,Adamczyk:2012ku,Back:2004zg,Doss:1987kq,Gutbrod:1989wd,HADES:2022osk}.

At the highest collision energies, at LHC and at RHIC, the elliptic flow at midrapidity is positive and is generated by the pressure  of the initial transverse overlap region defined by its spatial eccentricity, typically scaling via $v_n \sim \varepsilon_n \exp[\eta/s \cdot n^2]$, where $v_n$ is the n-th order flow coefficient, $\varepsilon_n$ is the n-th order eccentricity, $\eta$ is the shear viscosity and $s$ is the entropy density \cite{Shuryak:2013ke,Demir:2008tr}. Here, the system becomes deconfined and behaves as a nearly ideal liquid with small shear viscosity, $\eta$, to entropy density, $s$, ratio \cite{Ackermann:2000tr,Adler:2003cb,Huovinen:2001cy,Song:2007fn,Romatschke:2007mq,Luzum:2008cw}. This picture of the generation of elliptic flow starts to break around E$_\mathrm{lab} \approx 6A$ GeV where $v_2$ turns negative. In this energy regime the viscous corrections are large \cite{Demir:2008tr,Teslyk:2019ioo,Karpenko:2015xea,Rose:2017bjz,Reichert:2020oes,Hammelmann:2023fqw} and the spectators decouple on a timescale similar to the expansion of the fireball and thus partially interact with the compressed region \cite{OmanaKuttan:2023cno}. Therefore one is confronted with a highly intricate time evolution of the system. In the literature, there are two main, apparently competing, theories aiming to explain the negative $v_2$ at these energies:
\begin{enumerate}
    \item Squeeze-out \cite{LeFevre:2016vpp}: This mechanism describes that the density gradient from the initial overlap zone is stronger out-of-plane, where the density decreases from twice saturation density to the vacuum, than in-plane, where the density decreases from twice saturation density to saturation density. 
    The pressure then follows the density gradient and leads to enhanced out-of-plane emission. This means that $\langle p_y^2\rangle > \langle p_x^2\rangle$ due to the stronger out-of-plane acceleration.
    \item Shadowing \cite{Reichert:2023eev}: This mechanism suggests that the overlap eccentricity $\varepsilon_2$ leads initially, as in high energy collisions, to a stronger pressure in-plane than out-of-plane, although it is built up rather slowly. Particles with an in-plane momentum move then into the residual spectator located in coordinate space next to the central overlap region. These particles rescatter with spectator nucleons and loose in these collisions some of their in-plane momentum. Therefore $\langle p_x^2\rangle < \langle p_y^2\rangle$
due to rescattering with spectator nucleons.  At the same time the forward/backward rapidity of the spectator nucleons carries the $v_2$ from the pressure  to the forward/backward rapidity. 
\end{enumerate}

In the present article we aim to settle the question of the origin 
of the negative elliptic flow in the energy range of GSI/FAIR and RHIC-FXT. To study this we will employ numerical transport simulations using the Ultra-relativistic Quantum Molecular Dynamics approach (UrQMD). This allows to study the generation of elliptic flow in great detail, to quantify the contribution of potential interactions and collisions to the final measured flow coefficients. We will investigate semi-peripheral (20-30\% defined by a fixed impact parameter at b=7 fm) Au+Au collisions at E$_\mathrm{lab}=0.6A$ GeV, E$_\mathrm{lab}=1.23A$ GeV and $\sqrt{s_\mathrm{NN}}=3.0$ GeV, relevant for the HADES, STAR and upcoming CBM experiments. 

During the preparation of this article we noticed that a similar study has recently been done in \cite{Wang:2024ktk}, however focusing on lower energies and less sophisticated.

\section{Model set-up and flow extraction}

\subsection{UrQMD and the simulation set-up}
The Ultra-relativistic Quantum Molecular Dynamics (UrQMD) model \cite{Bass:1998ca,Bleicher:1999xi,Bleicher:2022kcu}, including all relevant hadrons and their resonances up to 4 GeV in mass, is employed in the present study. The UrQMD model is of QMD type \cite{Aichelin:1986wa,Aichelin:1991xy}. In a QMD model the nucleus wave function is a direct product of the nucleon wave functions $\Phi = \prod \phi_i$ where the wave function $\phi_i$ depends on two parameters ${\bf r}_i,{\bf p}_i$. The equations of motion of the nucleons can be derived from the Dirac-Frenkel-McLachlan variational principle \cite{Dirac_1930,Frenkel_1934,McLachlan:1964vaa,broeck:1988,raab:2000}.
When representing the nucleons as Gaussian wave functions $\phi_i$, the equations of motion of the parameters  ${\bf r}_i,{\bf p}_i$ become 
\begin{align}
    \dot{\mathbf{r}}_i = \frac{\partial \langle \mathcal{H} \rangle}{\partial \mathbf{p}_i} \quad \text{and} \quad \dot{\mathbf{p}}_i = \frac{\partial \langle \mathcal{H} \rangle}{\partial \mathbf{r}_i}
\end{align} 
with $\langle \mathcal{O} \rangle = \langle \Phi | \mathcal{O} | \Phi \rangle$ and $\mathcal{H} = T +  \mathcal{V} $ being the full n-body Hamiltonian, resembling the classical Hamilton's equations of motion, thus allowing to track individual nucleons through the time evolution. The potential employed is not the nucleon-nucleon potential but the real part of the Br\"uckner G-matrix. The imaginary part can be modeled by cross sections. They are taken either from experimental data, if available, or derived from effective model calculations. The excitation of resonances switches to a string picture at larger collision energies. 
In the present analysis we restrict the potential to a hard Skyrme type potential with parameters in accordance with \cite{Hillmann:2018nmd}, and neglect the discussion of momentum dependent potentials \cite{Aichelin:1987ti,Danielewicz:1999zn,Mohs:2024gyc,Steinheimer:2024eha,Kireyeu:2024hjo} and modified in-medium cross sections \cite{Li:2022wvu}.
\begin{align}
    U(\rho_\mathrm{B}) &= \alpha \left( \frac{\rho_\mathrm{B}}{\rho_0} \right) + \beta \left( \frac{\rho_\mathrm{B}}{\rho_0} \right)^\gamma
\end{align}
The UrQMD model has proven to reliably describe and predict many observables including flow in the presently investigated energy regime \cite{Hillmann:2019wlt,Reichert:2021ljd,Steinheimer:2022gqb} and is thus well suited for the current analysis. 

\subsection{Harmonic flow}
The final azimuthal momentum distribution of nucleons can be analyzed as
a Fourier series
\begin{equation}
    \frac{\mathrm{d}N(y,p_\mathrm{T})}{\mathrm{d}\phi} = 1 + 2 \sum\limits_{n=1}^\infty v_n(y,p_\mathrm{T}) \cos((\phi - \Psi_\mathrm{RP}))
\end{equation}
as first proposed in \cite{Voloshin:1994mz}\footnote{In the most general case an additional sine term appears as well \cite{Reichert:2022yxq}, however due to the point symmetry of the sine, its coefficients are zero when extracted with respect to the event planes and they also disappear on average when the reaction plane is used as reference angle.}. The $v_n$ are n-th order harmonic flow coefficients, $\tan(\phi) = p_y / p_x$ and $\Psi_\mathrm{RP}$ is the angle of the reaction plane which is fixed to $\Psi_\mathrm{RP}=0$ in the simulation. It should be pointed out that the reaction plane spanned by the impact parameter and the beam axis is unknown in experiments. To overcome this problem in high energy collision where the elliptic flow is positive, many methods have been developed, all having in common that the flow coefficients are not extracted with respect to the reaction plane, but with respect to the n-th order event plane \cite{Borghini:2002mv,Cheng:2000tk,Borghini:2001vi,Bhalerao:2003xf,Danielewicz:1985hn,Poskanzer:1998yz,Borghini:2000sa,Ollitrault:1997di}. 

In the presently investigated energy regime, the sign of the elliptic flow matters, and thus the measurement of $v_n$ with respect to the n-th order event plane is disfavored. In the negative-$v_2$ regime one has to utilize a good proxy for the reaction plane. Here, e.g. the HADES collaboration uses the first order event plane of the projectile spectators to approximate the event plane, allowing to measure the sign of the elliptic flow \cite{Kardan:2017knj,HADES:2020lob,HADES:2022osk} while the STAR collaboration uses their event plane detector to estimate the first order event plane \cite{STAR:2021ozh,STAR:2021yiu}.

While measurements and calculations of higher order flow harmonics are valuable in itself \cite{Alver:2010gr,Schenke:2010rr,Petersen:2010cw,Hillmann:2019wlt}, in this study we focus on the directed flow $v_1$ and the elliptic flow $v_2$. Both can be expressed either in terms of the azimuthal angle or equivalently by the momenta in the transverse plane and are calculated via
\begin{align}
    \langle v_1 \rangle &= \langle \cos(\phi) \rangle = \bigg\langle \frac{p_\mathrm{x}}{p_\mathrm{T}} \bigg\rangle \\
    \langle v_2 \rangle &= \langle \cos(2 \phi) \rangle = \bigg\langle \frac{p_\mathrm{x}^2-p_\mathrm{y}^2}{p_\mathrm{T}^2} \bigg\rangle
\end{align}
in which $p_\mathrm{T}$ is the transverse momentum and the average $\langle \cdot \rangle$ runs over particles in a given ensemble and phase space window. From this definition their interpretation and symmetries are also evident. At certain passages in this article we will make use of the rapidity-signed directed flow $v_1^*$ defined as
\begin{equation}
    v_1^* = sgn(y) \cdot v_1
\end{equation}
where $sgn(\cdot)$ is the sign function and $y$ is the rapidity. The quantity $v_1^*$ will be referred to as \textit{signed directed flow}.

\section{Results}
We start the search for the origin of the directed and elliptic flow by analyzing the time dependence of $v_1$ and $v_2$. The time dependent growth of both observables will shed light on their generation mechanism. To understand the development of flow, the different timescales at the different collision energies have to be considered. The timescale of the whole evolution is set by the geometric time of full overlap, which for a symmetric collision system is given by
\begin{align}\label{eq:toverlap}
    t_\mathrm{overlap} &= \frac{R}{\gamma\beta} = \frac{2 m_p R_0 \sqrt[3]{A}}{\sqrt{\sqrt{s_\mathrm{NN}}^2 - 4m_p^2}}.
\end{align}
It can be expressed by system quantities using $R=R_0\sqrt[3]{A}$ ($R_0^\mathrm{UrQMD}=1.21$ fm), $\gamma = \sqrt{s_\mathrm{NN}}/(2m_p)$ and $\gamma^{-1} = \sqrt{1-\beta^2}$. It should be noted beforehand that the geometric time of full overlap (neglecting compression) in any collision is a lower boundary/estimate for the realized time of full overlap or maximal compression due to the deceleration of the impinging nuclei. At $1.23A$ GeV kinetic beam energy $t_\mathrm{overlap} = 8.7$ fm/c, however, keeping in mind the compression of the system, one can expect the realistic time of full overlap to be a bit later, $t_\mathrm{overlap} \approx 10-12$ fm/c lining up with the time of maximal compression obtained in previous studies \cite{OmanaKuttan:2022the,Reichert:2020oes}. At the other investigated collision energies the geometric overlap times are: $t_\mathrm{overlap}^{\mathrm{E}_\mathrm{lab}=0.6A\,\mathrm{GeV}}=10.8$ fm/c and $t_\mathrm{overlap}^{\sqrt{s_\mathrm{NN}}=3.0\,\mathrm{GeV}}=5.6$ fm/c.

\subsection{Time evolution of elliptic and directed flow in the x-z and x-y plane}
\begin{figure*} [t!hb]
    \centering
    \includegraphics[width=\columnwidth]{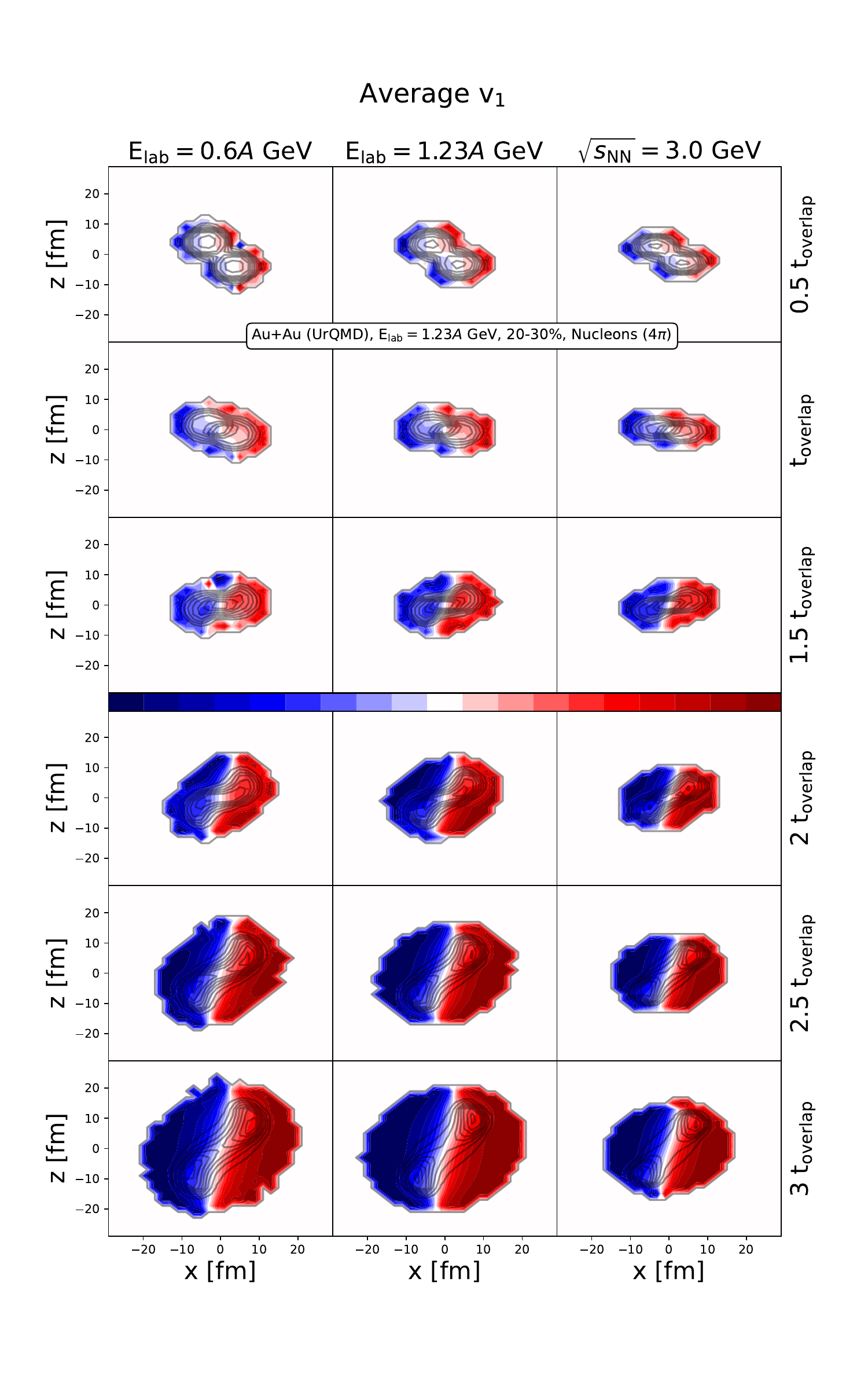}
    \includegraphics[width=\columnwidth]{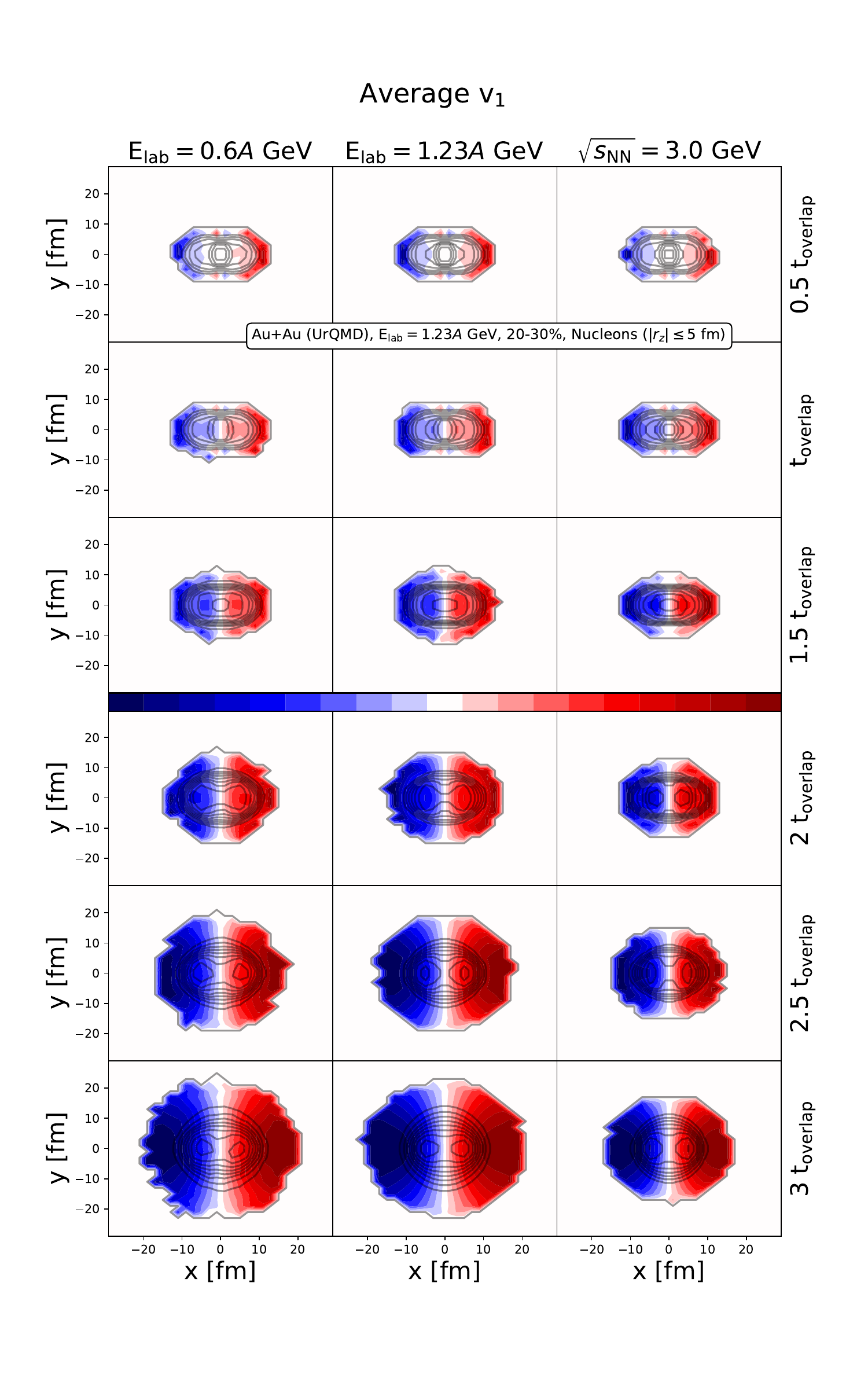}
    \caption{[Color online] Both panels show the time evolution of the directed flow $v_1$. The left panel shows $v_1$ of baryons in the x-z plane in $4\pi$ at time $t$ while the right panel shows $v_1$ of baryons in the x-y plane with $|r_z|\leq5$ fm at time t. In each panel the left column shows results at $0.6A$ GeV kinetic beam energy, the central column at $1.23A$ GeV kinetic beam energy and the right column shows results at $3.0$ GeV center of mass energy. The time is denoted on the right axis and is scaled by the energy dependent time of full overlap. The color bar in the middle of the figure encodes the magnitude of $v_1$ (from -1 to 1, centered at zero). The black contour lines denote the number density. All panels display Au+Au collisions at b=7 fm calculated with UrQMD with a hard Skyrme EoS.}
    \label{fig:v1_xz_xy}
\end{figure*}

\begin{figure*} [t!hb]
    \centering
    \includegraphics[width=\columnwidth]{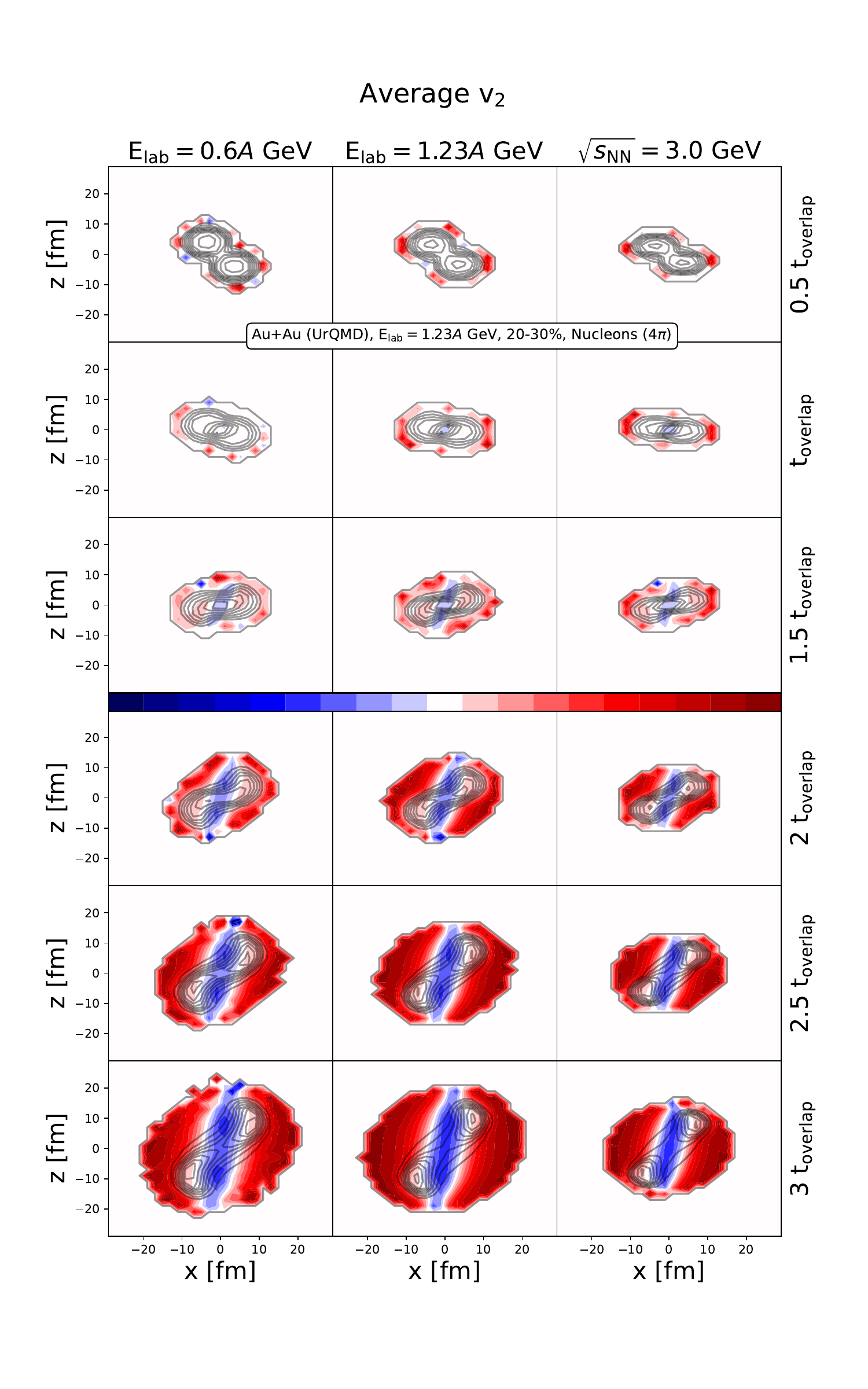}
    \includegraphics[width=\columnwidth]{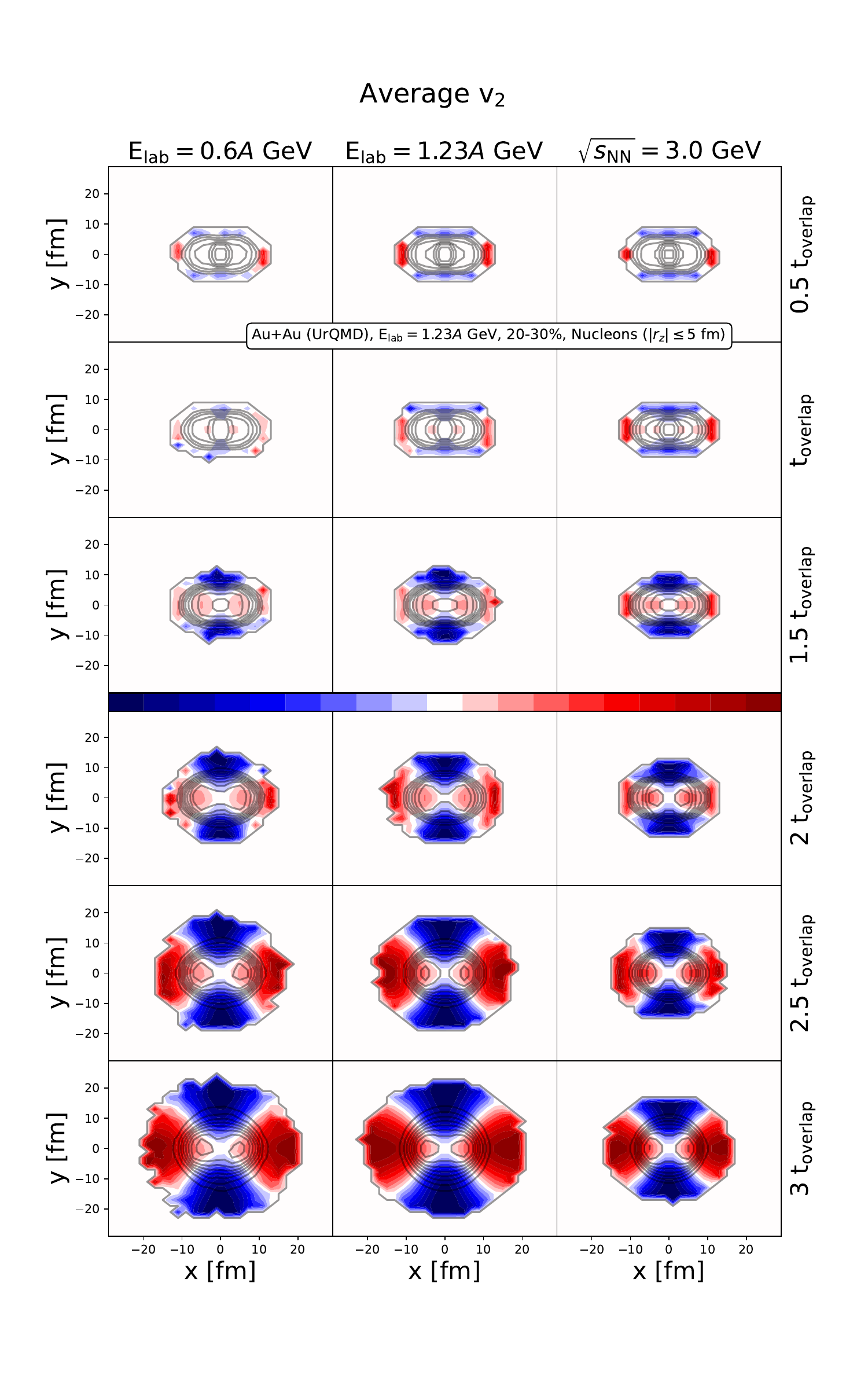}
    \caption{[Color online] Both panels show the time evolution of the elliptic flow $v_2$. The left panel shows $v_2$ of baryons in the x-z plane in $4\pi$ at time $t$ while the right panel shows $v_2$ of baryons in the x-y plane with $|r_z|\leq5$ fm at time t. In each panel the left column shows results at $0.6A$ GeV kinetic beam energy, the central column at $1.23A$ GeV kinetic beam energy and the right column shows results at $3.0$ GeV center of mass energy. The time is denoted on the right axis and is scaled by the energy dependent time of full overlap. The colored bar in the middle of the figure encodes the magnitude of $v_2$ (from -1 to 1, centered at zero). The black contour lines denote the number density. All panels display Au+Au collisions at b=7 fm calculated with UrQMD with a hard Skyrme EoS.}
    \label{fig:v2_xz_xy}
\end{figure*}
First we will investigate the time evolution of the directed flow $v_1$ and of the elliptic flow $v_2$ in the x-z and x-y planes. In order to take into account the different time scales at the different energies, we have scaled the time evolution by $t_\mathrm{overlap}$ as given in Eq. \eqref{eq:toverlap}, and show the flow evolution at the times: 0.5, 1, 1.5, 2, 2.5 and 3 times $t_\mathrm{overlap}$. In order to calculate the local flow coefficients we will put a grid onto the x-y and x-z planes with $\mathrm{d}r=2$ fm and calculate the average $\langle v_1 \rangle$ and $\langle v_2 \rangle$ in each cell of the grid, where $\langle\cdot\rangle$ indicates taking the average per particle in the respective cell, and present the results in Sec. \ref{sec:flow_plane}. 
This analysis allows to identify the regions in the system with the strongest magnitude of the flow coefficients but not their weight to the total flow in the (average) event because it is not weighted by the total number of nucleons present at this space time point.
Therefore, in a second step we will calculate in Sec. \ref{sec:weighted_flow_plane} the flow coefficients $\langle v_1 \rangle$ and $\langle v_2 \rangle$ weighted with the local density $\rho$ in each cell of the grid. This allows for better identification of regions in system evolution that actually contribute to the integrated value of the directed or elliptic flow in the respective time step.

\subsubsection{Time evolution of the average directed and elliptic flow}\label{sec:flow_plane}
Figs. \ref{fig:v1_xz_xy} and \ref{fig:v2_xz_xy} show the time evolution of $\langle v_1 \rangle$ and $\langle v_2 \rangle$, respectively, in the x-z and x-y planes. Here the average $\langle v_1 \rangle$ and $\langle v_2 \rangle$ are averaged over all baryons in each cell of the grid. In both figures the left-hand panel shows $v_1$ ($v_2$) in the x-z plane in $4\pi$ while the right-hand panel shows $v_1$ ($v_2$) in the x-y plane for $|r_z|\leq5$ fm. The average nucleon flow densities are shown at the fixed times: 0.5, 1, 1.5, 2, 2.5 and 3 times $t_\mathrm{overlap}$ (as denoted on the right hand side of each plot). In each panel the three columns show the time evolution at three collision energies, i.e. at $0.6A$ GeV kinetic beam energy (left), $1.23A$ GeV kinetic beam energy (middle) and $3.0$ GeV center-of-mass energy (right). The colored bar encodes the magnitude of $v_1$ ($v_2$) at each space-time point and varies from -1 to 1, being centered around 0. In addition, the density of the nuclei and the fireball is shown as black contour lines.

Before analyzing the figures in detail, the first major observation is that the flow and density evolution in the x-z and x-y plane behave very similar (up to contraction in z-direction due to the Lorentz $\gamma$ factor) at the three different collision energies, both for the directed flow and the elliptic flow, if the time is scaled by the geometric time of full overlap.

We will now systematically investigate the time evolution. At $t=0.5t_\mathrm{overlap}$ $v_1$ shows mostly zero directed flow with a small exception at the outwards facing surface of the nuclei, caused by the kinematics in one single nucleus and only little influenced by the heavy-ion collision. At the same time also the elliptic flow is mostly zero. The colored regions reflect the movement of the nucleons in the potential of the other nucleons of the same nucleus and are not due to the heavy-ion collision with the exception of the negative $v_2$ value at $r_x=0$ and $|r_y| \approx R$. There one can already observe a region of negative $v_2$ forming at the boundary between the overlap region and the vacuum at small $|r_x|$ and large $|r_y|$, i.e. at the ``tips of the almond shape''. The contour lines reveal that also the density gradient is very strong in this region, supporting that the pressure is initially stronger out-of-plane than in-plane, as suggested in \cite{LeFevre:2016vpp}. However, comparing the x-y with the x-z planes, it is also evident that the two impinging nuclei have barely touched at all energies. 
\begin{figure*} [t!hb]
    \centering
    \includegraphics[width=\columnwidth]{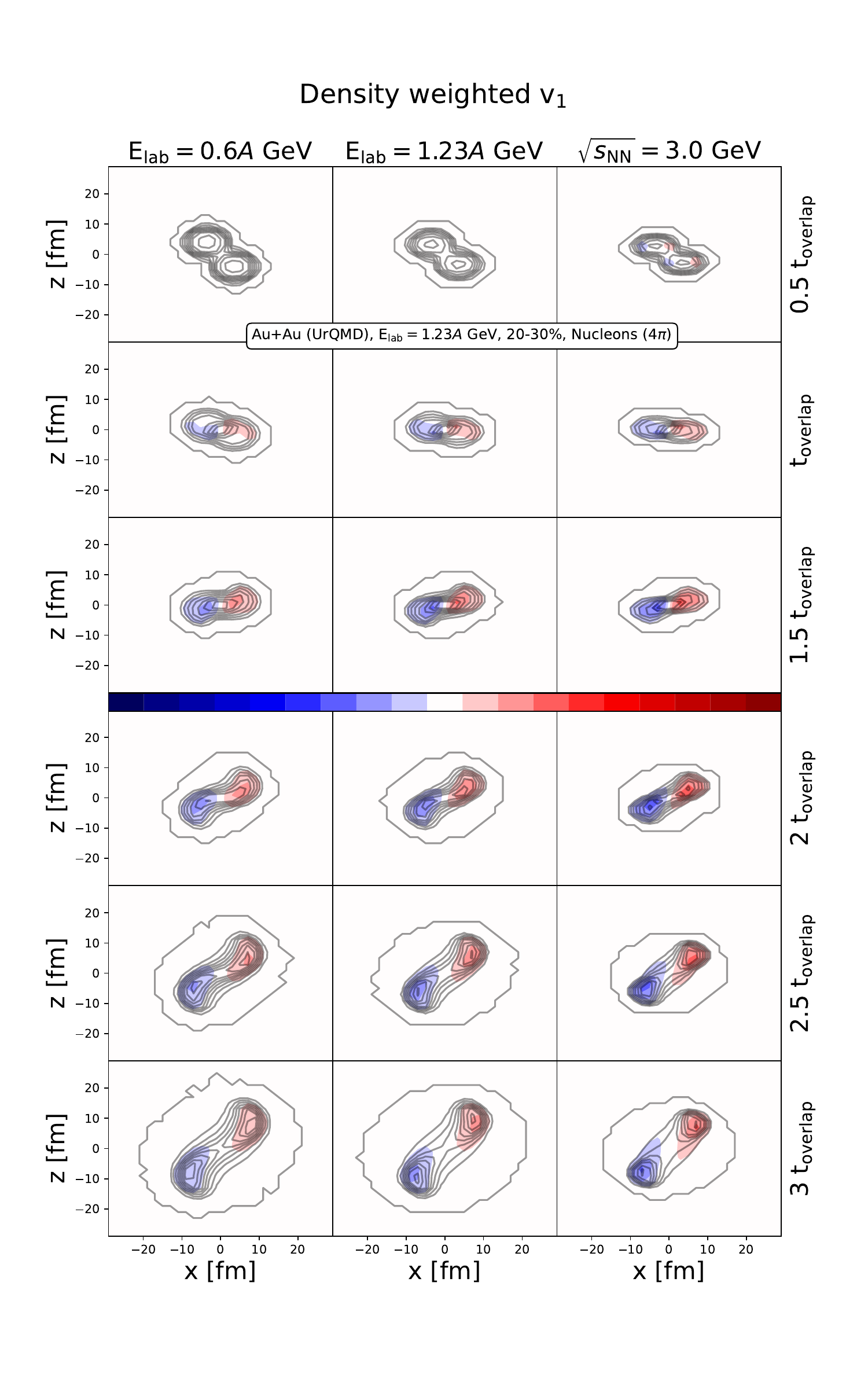}
    \includegraphics[width=\columnwidth]{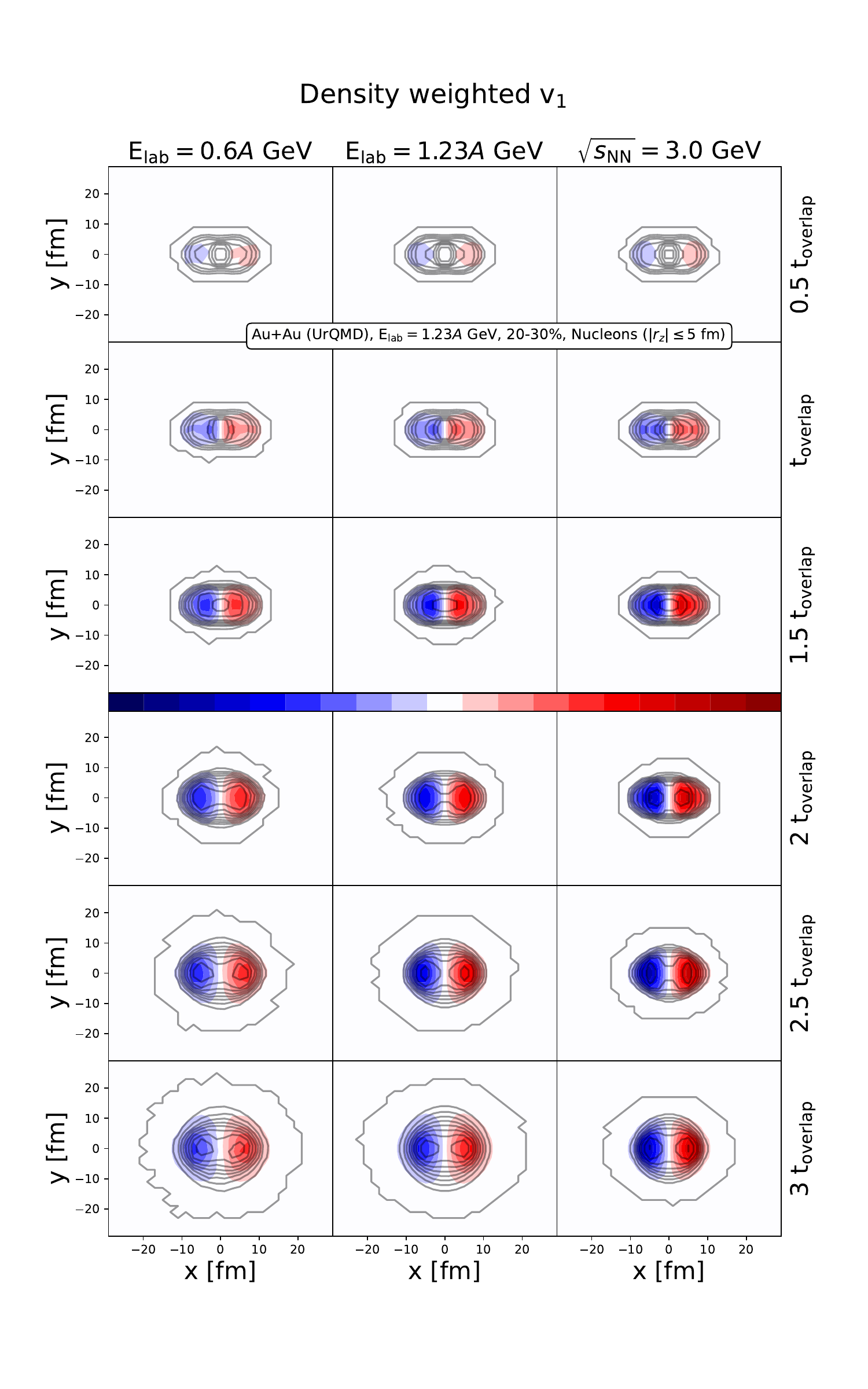}
    \caption{[Color online] Both panels show the time evolution of the density weighted directed flow $v_1$. The left panel shows the density weighted $v_1$ of baryons in the x-z plane in $4\pi$ at time $t$ while the right panel shows the density weighted $v_1$ of baryons in the x-y plane with $|r_z|\leq5$ fm at time t. In each panel the left column shows results at $0.6A$ GeV kinetic beam energy, the central column at $1.23A$ GeV kinetic beam energy and the right column shows results at $3.0$ GeV center of mass energy. The time is denoted on the right axis and is scaled by the energy dependent time of full overlap. The colored bar in the middle of the figure encodes the magnitude of the density weighted $v_1$ (symmetric, centered at zero). The black contour lines denote the number density. All panels display Au+Au collisions at b=7 fm calculated with UrQMD with a hard Skyrme EoS.}
    \label{fig:v1_xz_xy_weighted}
\end{figure*}

\begin{figure*} [t!hb]
    \centering
    \includegraphics[width=\columnwidth]{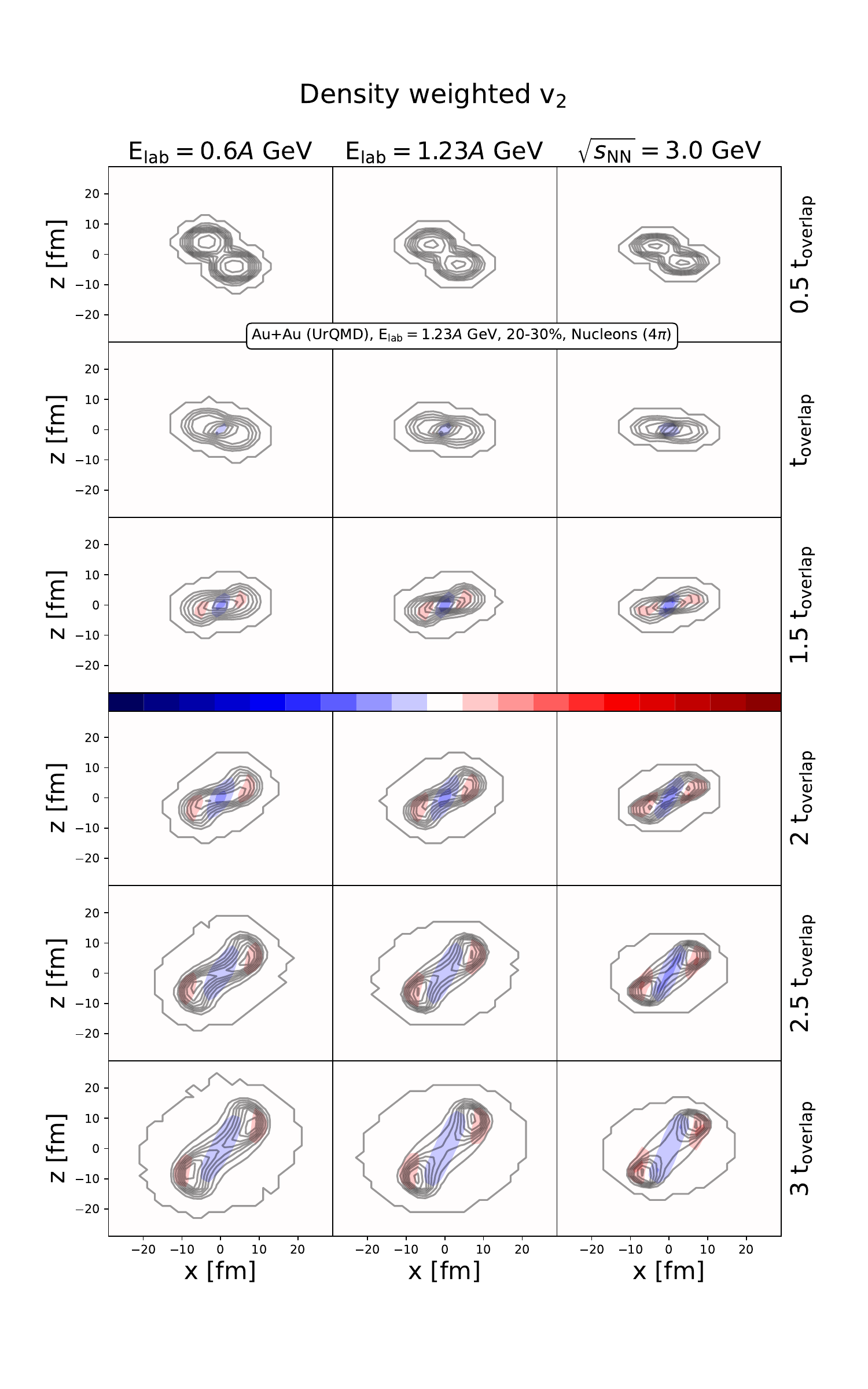}
    \includegraphics[width=\columnwidth]{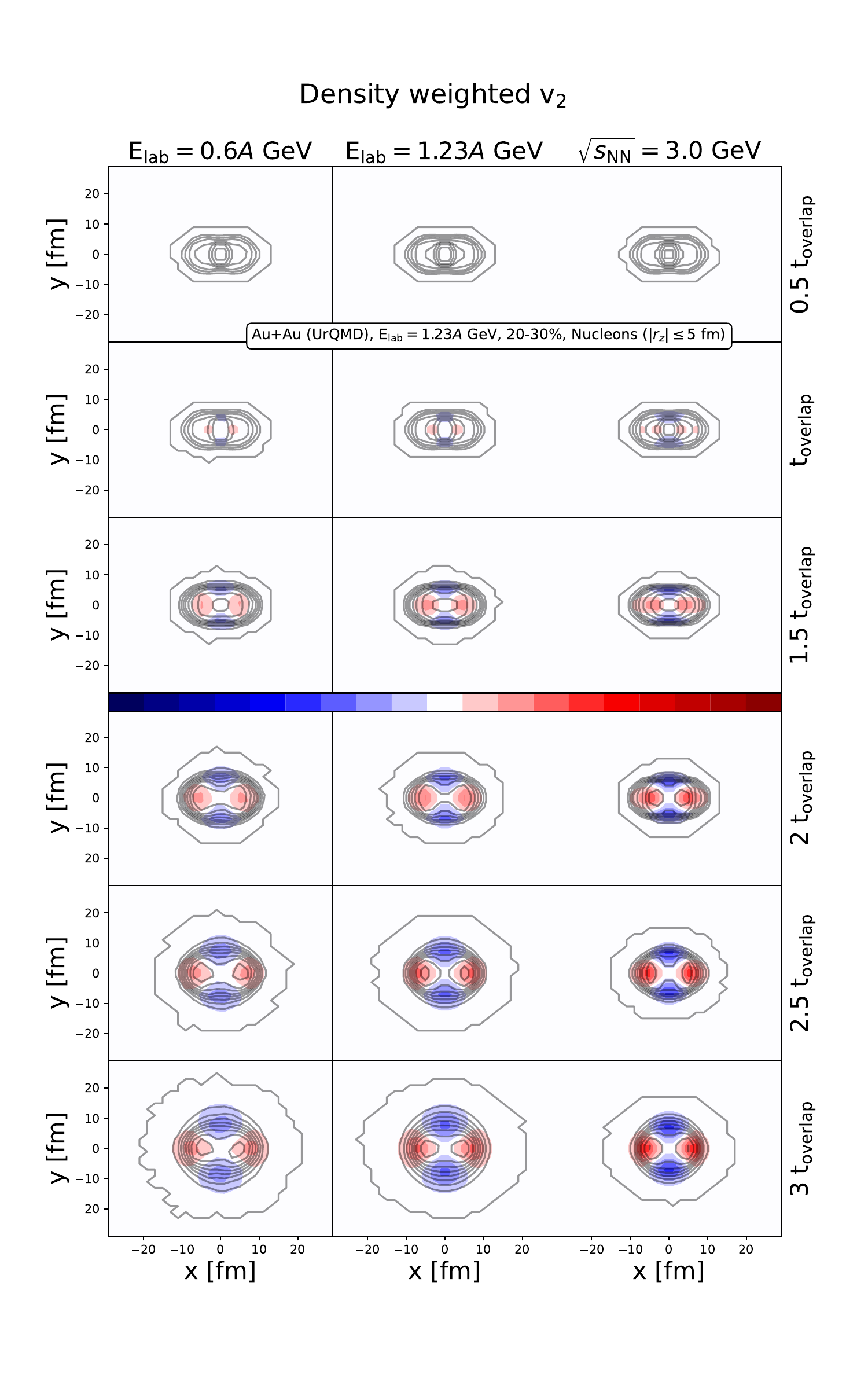}
    \caption{[Color online] Both panels show the time evolution of the density weighted elliptic flow $v_2$. The left panel shows the density weighted $v_2$ of baryons in the x-z plane in $4\pi$ at time $t$ while the right panel shows the density weighted $v_2$ of baryons in the x-y plane with $|r_z|\leq5$ fm at time t. In each panel the left column shows results at $0.6A$ GeV kinetic beam energy, the central column at $1.23A$ GeV kinetic beam energy and the right column results at $3.0$ GeV center of mass energy. The time is denoted on the right axis and is scaled by the energy dependent time of full overlap. The colored bar in the middle of the figure encodes the magnitude of the density weighted $v_2$ (symmetric, centered at zero). The black contour lines denote the number density. All panels display Au+Au collisions at b=7 fm calculated with UrQMD with a hard Skyrme EoS.}
    \label{fig:v2_xz_xy_weighted}
\end{figure*}

Moving forward to the time of full overlap $t=t_\mathrm{overlap}$, 
we observe  that $v_1$ starts to develop in the range of the potential around the overlap volume.

At this time the two nuclei are nearly next to each other\footnote{Due to the deceleration of baryons, the two nuclei will be exactly next to each other slightly later.} and one can observe the formation of a negative $v_2$ region in the center of the system in the x-z plane, as indicated by the bright bluish region. By comparison to the x-y plane at full overlap, this region is still located at the tips of the almond shaped overlap region and it is mostly consisting of high $p_\mathrm{T}$ particles. However, in the x-y plane one also observes a small region with positive $v_2$ forming at the interface between the stopped fireball and the bypassing spectator matter. 

At $t=1.5t_\mathrm{overlap}$, coinciding with the time when the  maximal density and compression has been obtained in previous studies with UrQMD at similar energies \cite{OmanaKuttan:2022the,Reichert:2020oes}, the two impinging nuclei have passed each other half way through.
At this time the size of the system in transverse direction has not increased substantially. Now $v_1$ is fully developed
and pushes the two hemispheres away from each other. 

One observes also that the negative $v_2$ region is growing in the x-z plane and its symmetry axis becomes tilted towards the direction of the deflection of the two nuclei, seen in the $v_1$ figure. The rest of the system is showing a positive $v_2$ in the x-z plane. In the x-y plane the regions with strongly negative $v_2$ at the tips of the overlap zone have expanded further into the vacuum, reducing the strong gradient in out-of-plane direction. At the same time the density in the center of the system has increased drastically reaching 2-4 times saturation density and the pressure in the in-plane direction is becoming more dominant in the center of the system as seen by the contour lines and the red color shading. By comparison to $v_1$ in the x-z and x-y planes one notices that the directed flow is separated by the same tilted axis defined by the region of negative elliptic flow in the x-z plane, while in the x-y plane the directed flow is progressively becoming stronger. Due to the stopping and deceleration in the overlap zone and because the residual nuclei have only partially passed each other, roughly half of the typical ``spectator'' is still located next to the overlap region in coordinate space. It did not decelerate substantially and is thus still located at forward/backward rapidity in momentum space.

From two to three times the time of full overlap the system starts to expand and hard collisions become less frequent. Therefore coordinate space-momentum space correlations become more important. At large $r_x$ and $r_y$ values we can only find nucleons which have a large momentum in these directions. This by itself defines regions of large $v_2$.
We summarize the findings of these time steps. In the late stage of the evolution we see mostly coordinate space-momentum space correlations where in the x-y plane at $45^\circ$ the $v_2$ changes from positive to negative values. Only in the coordinate space region
where the nuclei have originally been located we can still observe how the $v_2$ is influenced by collisions and the potential.
There we see a bridge, which connects projectile and target,
and which is dominated by a negative $v_2$, even if one sees from the x-y plot that also a (subdominant) positive $v_2$ component is present. This positive component increases
with beam energy simultaneously with the increasing eccentricity of the overlap region. The positive $v_2$ value observed there is therefore a precursor of the hydrodynamical flow caused by the eccentricity, which has been observed for much higher beam energies \cite{Shuryak:2013ke,Demir:2008tr,Ackermann:2000tr,Adler:2003cb,Huovinen:2001cy,Song:2007fn,Romatschke:2007mq,Luzum:2008cw}.
Together this reminds on a dominant squeeze-out accompanied by a subdominant in-plane flow due to compression.

\subsection{Time evolution of the density weighted elliptic and directed flow}\label{sec:weighted_flow_plane}
We have now systematically investigated the spatial regions of strongly positive and negative directed and elliptic flow during the time evolution of semi-peripheral Au+Au collisions at three different energies. This allows to identify where the nucleons, having the largest $v_1$ and $v_2$ values, are localized during the heavy-ion collision. This presentation does, however, not allow to evaluate their contribution to the space integrated $v_1$ and $v_2$, which are measured experimentally.

To obtain the integrated harmonic flow coefficients $v_1$ and $v_2$, which can be compared with experimental results, we have to weight the local flow coefficients with the number of baryons which are in the same cell in coordinate space normalized to all baryons present in the system. 

Figs. \ref{fig:v1_xz_xy_weighted} and \ref{fig:v2_xz_xy_weighted} show the time evolution of the density weighted directed flow $v_1$ and the density weighted elliptic flow $v_2$, respectively. In both figures, the left-hand panel shows the density weighted $v_1$ ($v_2$) in the x-z plane in $4\pi$ while the right-hand panel shows the density weighted $v_1$ ($v_2$) in the x-y plane at $|r_z|\leq5$ fm. The densities are shown at the fixed times: 0.5, 1, 1.5, 2, 2.5 and 3 times $t_\mathrm{overlap}$ (as denoted on the right hand side of each plot). In each panel the three columns show the time evolution at three collision energies, i.e. at $0.6A$ GeV kinetic beam energy (left), $1.23A$ GeV kinetic beam energy (middle) and $3.0$ GeV center-of-mass energy (right). The colored bar denotes the magnitude of the density weighted $v_1$ ($v_2$) at each space-time point and is symmetric and centered around 0. In addition, the profile of the nuclear density is shown as black contour lines.

The first observation, in comparison to the previously discussed calculations, is that the regions far away from the center of the fireball do not contribute to the measured flow due to their comparatively small particle number density. The time evolution at the three different energies is again behaving very similar, for the directed as well as for the elliptic flow.
\begin{figure} [t!hb]
    \centering
    \includegraphics[width=\columnwidth]{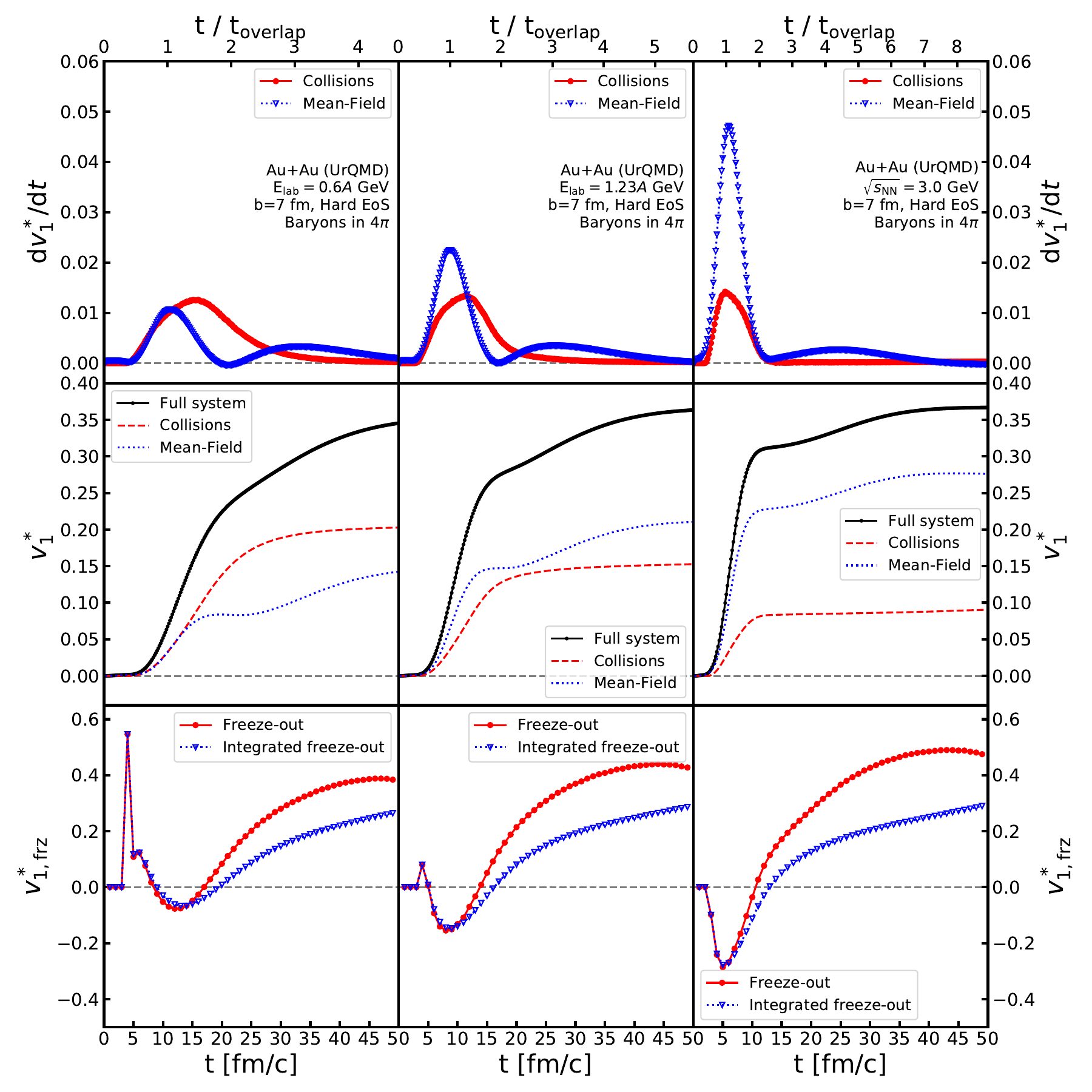}
    \caption{[Color online] Time dependence of $v_1^*$ of all baryons at in $4\pi$ in 20-30\% peripheral Au+Au collisions calculated with UrQMD. From left to right the columns show calculations at E$_\mathrm{lab}=0.6A$ GeV, E$_\mathrm{lab}=1.23A$ GeV and $\sqrt{s_\mathrm{NN}}=3.0$ GeV. The first row shows the differential change of the directed flow with time $\mathrm{d}v_1^*/\mathrm{d}t$ by collisions (red circles) and by the potential (blue triangles). The second row shows the directed flow of the whole system at time $t$ (black line) as well as the integrated differential change by collisions (red) and by the potential (blue). The last row shows the directed flow of nucleons freezing out at time $t$ (red) and the directed flow of all nucleons that have frozen out until time $t$ (blue).}
    \label{fig:v1_4pi_time}
\end{figure}

For the directed flow one sees in the x-z plane as well as in the x-y plane what one expects if $v_1$ is produced by the density gradient of the potential, although we will see in the next section that in reality the situation is more complex. It is remarkable that in a large area around $r_x\approx r_y\approx 0$ the directed flow is zero. Clearly at the end of the reaction $v_1$ is for all energies most prominent at the outer edge of the spectators. Thus $v_1$ is communicated by interactions to all spectator nucleons. Between the three energies we find only quantitative differences.

Starting at $t=0.5t_\mathrm{overlap}$ one notes that there is no apparent elliptic flow yet, only the directed flow displays a tiny deviation from zero in the x-y plane, denoted by the faded colors. In comparison to the previous discussion of flow without density weight, the region with initial strongly negative $v_2$ at the tips of the overlap region vanished. Although the out-of-plane pressure is very strong, the number of particles being accelerated in the gradient is small at early times. 

At the time of full geometric overlap $t=t_\mathrm{overlap}$ the two nuclei are again nearly next to each other and the overlapping, decelerating matter in the center of the system develops a negative $v_2$ in the x-z plane. In the x-y plane one observes that it is composed of two contributions: one observes a negative elliptic flow in out-of-plane (at the tips) as well as a positive elliptic flow in in-plane direction (at the sides of the almond shaped overlap region). Integrating over the y-axis here leads to an overall negative $v_2$, as seen in the x-z plane. It is  worthwhile noting that the density weighted $v_2$ shows that the gradient out-of-plane and the gradient in-plane are acting at the same time and compete with each other. 

The density weighted directed flow is getting stronger during the same time step and is present left and right to the center stretching into the spectating nucleons. Because the density profile (denoted by the black contour lines) is not changing a lot, the $v_2$ out-of-plane and in-plane are caused by an acceleration due to the potential gradient in y- and x-direction, respectively. Nucleons located directly in the center of the system do not acquire a $v_2$ due to symmetry reasons.

As previously mentioned, at the time of maximal compression, at $t=1.5t_\mathrm{overlap}$,  
the size of the system in transverse direction has not yet changed,
therefore coordinate space-momentum space correlations are not yet present.
\begin{figure} [t!hb]
    \centering
    \includegraphics[width=\columnwidth]{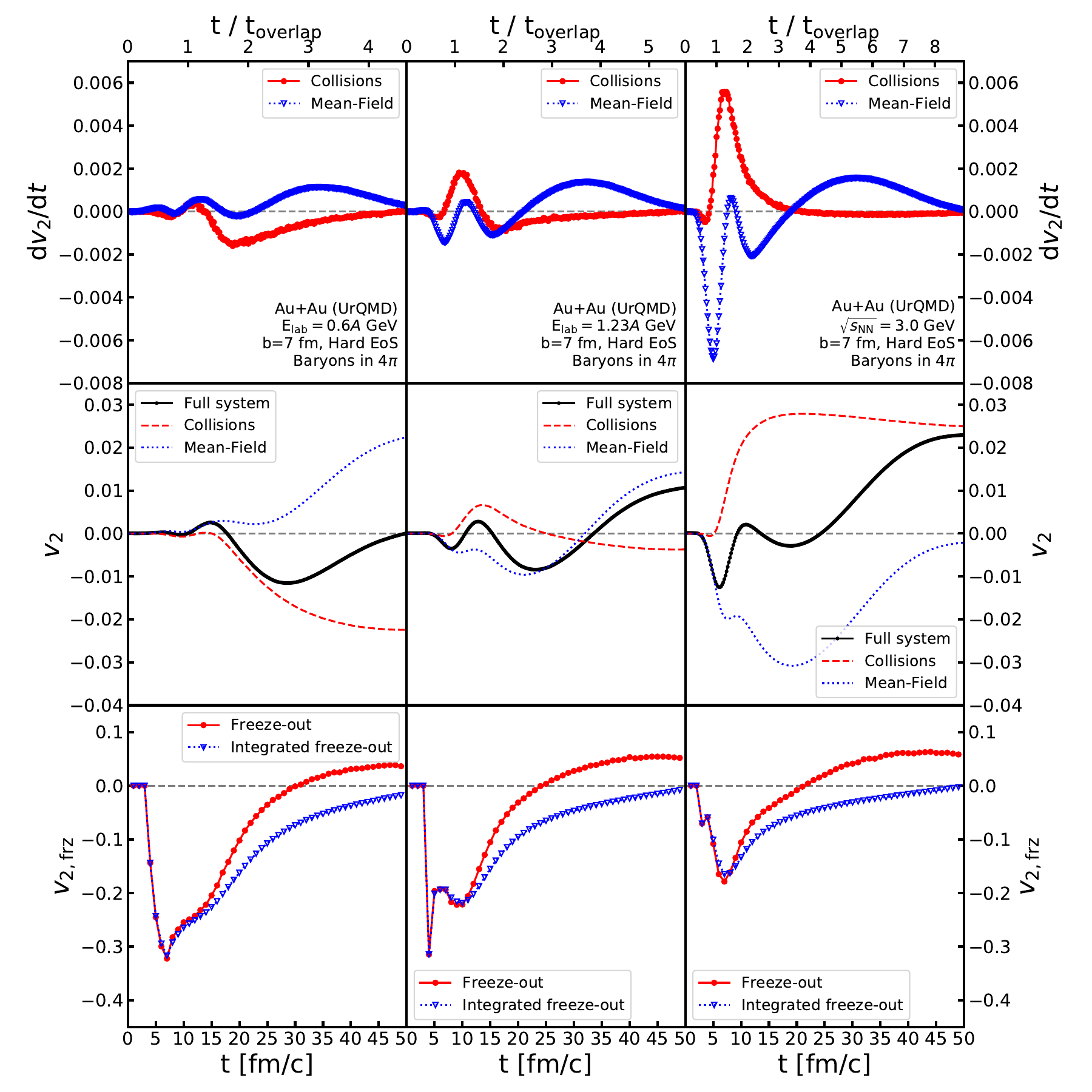}
    \caption{[Color online] Time dependence of $v_2$ of all baryons at in $4\pi$ in 20-30\% peripheral Au+Au collisions calculated with UrQMD. From left to right the columns show calculations at E$_\mathrm{lab}=0.6A$ GeV, E$_\mathrm{lab}=1.23A$ GeV and $\sqrt{s_\mathrm{NN}}=3.0$ GeV. The first row shows the differential change of the elliptic flow with time $\mathrm{d}v_2/\mathrm{d}t$ by collisions (red circles) and by the potential (blue triangles). The second row shows the elliptic flow of the whole system at time $t$ (black line) as well as the integrated differential change by collisions (red) and by the potential (blue). The last row shows the elliptic flow of nucleons freezing out at time $t$ (red) and the elliptic flow of all nucleons that have frozen out until time $t$ (blue).}
    \label{fig:v2_4pi_time}
\end{figure}
We see a region with positive $v_2$ showing up in the x-z plane at the surface between the overlap zone and the bypassing matter. Again, the shapes are very similar at all three energies. Moving to the x-y plane, one observes that both, the regions of positive and negative elliptic flow, are growing with energy. Although the area of negative $v_2$ is smaller compared to that showing positive elliptic flow, its magnitude is stronger in the out-of-plane direction, with the consequence that the total integrated $v_2$ is negative. 
In the subsequent time steps the positive $v_2$ areas penetrate into the spectator matter until they arrive at the outer edge and are transported towards areas with a strong positive resp. negative rapidity. The area with a negative $v_2$ remains centered at midrapidity but extends further when the matter bridge between projectile and target gets more and more stretched. At the end at $z\approx 0$ (which corresponds roughly to the contribution of midrapidity particles)  $v_2$ is negative. In the x-y projection we see that the positive and negative $v_2$ regions are separated in transverse direction. Although there is still matter around $r_x\approx r_y\approx 0$ it does, due to symmetry reasons,  not contribute to $v_2$.
\begin{figure} [t!hb]
    \centering
    \includegraphics[width=\columnwidth]{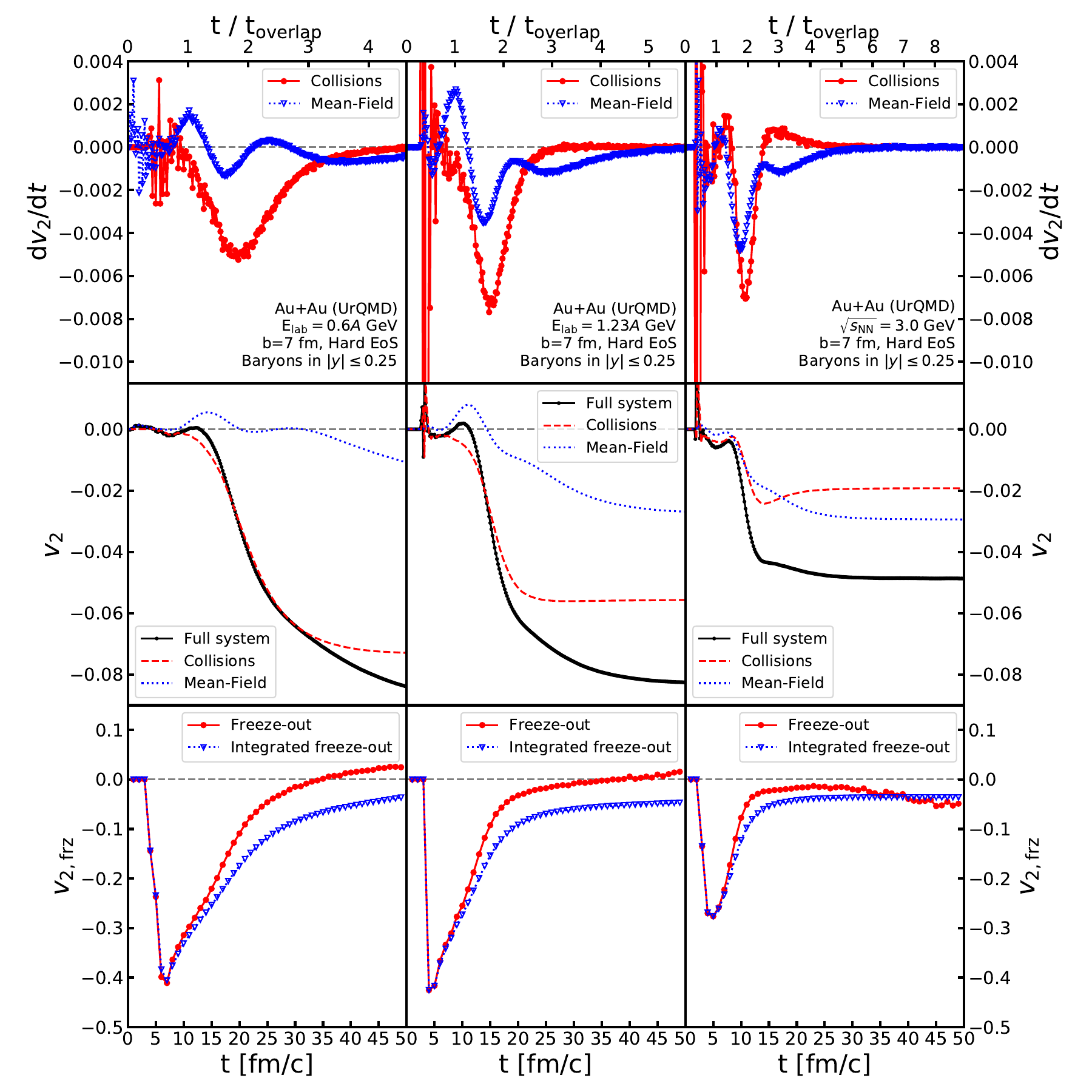}
    \caption{[Color online] Time dependence of $v_2$ of all baryons at midrapidity $|y|\leq0.25$ in 20-30\% peripheral Au+Au collisions calculated with UrQMD. From left to right the columns show calculations at E$_\mathrm{lab}=0.6A$ GeV, E$_\mathrm{lab}=1.23A$ GeV and $\sqrt{s_\mathrm{NN}}=3.0$ GeV. The first row shows the differential change of the elliptic flow with time $\mathrm{d}v_2/\mathrm{d}t$ by collisions (red circles) and by the potential (blue triangles). The second row shows the elliptic flow of the whole system at time $t$ (black line) as well as the integrated differential change by collisions (red) and by the potential (blue). The last row shows the elliptic flow of nucleons freezing out at time $t$ (red) and the elliptic flow of all nucleons that have frozen out until time $t$ (blue).}
    \label{fig:v2_ymid_time}
\end{figure}

\subsection{Quantitative assessment of the influence of the EoS and scatterings}
So far we have presented the average and the density weighted directed and elliptic flow in the x-z plane and the x-y plane. This has revealed that especially the elliptic flow has a complicated origin. There exist regions of negative $v_2$ and regions with positive $v_2$ whose origin as well as whose time evolution is different.

As a next step, we want to condense our calculations to make them comparable with the experimental observables by integrating over  coordinate space. To study the origin of the flow coefficient we  differentiate furthermore between the contribution of collisions and of the potential interaction.
Here we take advantage of the QMD character of the UrQMD model where the time evolution equations are solved in a finite time step ($\delta t$) method and where it is possible to track individual hadrons during the time evolution. We define $\mathrm{d}v_n/\mathrm{d}t$ as the differential change of the n-th order harmonic flow between two time steps\footnote{When employing a QMD type simulation with potentials, the time is propagating in fixed small time steps on the order of $\delta t=0.2$ fm/c. See \cite{Bass:1998ca,Bleicher:1999xi} for more details.} during the evolution of the system, i.e. $\mathrm{d}v_n/\mathrm{d}t(t) = (v_n(t) - v_n(t-\delta t)) / \delta t$. The set up of UrQMD allows to separate $\mathrm{d}v_n/\mathrm{d}t$ from collisions and from the potential interactions within a time step. Due to momentum conservation $v_1$ integrated over coordinate space will be zero, hence in this section we will make use of the signed directed flow $v_1^* = sgn(y) \cdot v_1$.

Figs. \ref{fig:v1_4pi_time}, \ref{fig:v2_4pi_time} and \ref{fig:v2_ymid_time} show the time evolution of the signed directed flow coefficient $v_1^*$ in $4\pi$, the elliptic flow $v_2$ in $4\pi$ and the elliptic flow at midrapidity $|y|\leq0.25$, respectively. Each figure shows the flow coefficients at the beam energies of $0.6A$ GeV (first column) and of $1.23A$ GeV (second column) as well as for $3.0$ GeV center-of-mass energy (third column). The time is denoted on the bottom x-axis in units of fm/c and on the top x-axis it is scaled by the time of full overlap at the respective energy. The first row shows the differential change of the flow with time $\mathrm{d}v_n/\mathrm{d}t$ by collisions (red circles) and by the potential (blue triangles). The second row shows the (up to t integrated) flow of the whole system at time $t$ (black line) as well as the integrated differential change by collisions (red) and by the potential (blue). The last row shows the flow of nucleons freezing out at time $t$ (red) and the time integrated flow of all nucleons that have frozen out until time $t$ (blue). All plots depict Au+Au collisions at a fixed impact parameter, $b=7$ fm, simulated by UrQMD with a hard Skyrme type EoS. For an easy comparison the limits on the y-axis are equal in each row throughout Figs. \ref{fig:v1_4pi_time}, \ref{fig:v2_4pi_time} and \ref{fig:v2_ymid_time}. 


\subsubsection{Directed flow $v_1$ in $4\pi$}\label{Sec:flow_4pi}
We start out with the discussion of the signed directed flow coefficient in $4\pi$.
For the full phase space acceptance the signed directed flow $v_1$ of baryons starts to deviate from zero as soon as the nucleons feel the potential created by the nucleons of the other nucleus (at roughly $t=0.5 \tmo$) and is rapidly growing. 

For $v_1^*$ the general features for the further time evolution are similar at all energies. What differs is the relative contribution of the potential and collisions and the time point when they contribute. The $\mathrm{d}v_1^*/\mathrm{d}t$ generated by the potential peaks for all energies at the time of full overlap and increases with energy due to the increasing compression of the overlap zone, while the change through collisions peaks later at the lower energies but roughly at the same time for $\sqrt{s_\mathrm{NN}}=3.0$ GeV. The importance of the potential contribution increases with energy whereas that of collisions decreases. There is a second peak of the potential contribution around $2-3\tmo$, which contributes roughly 20\% to the final value of $v_1$ at the two lower energies. We will come back to this observation later.


\subsubsection{Elliptic flow $v_2$ in 4$\pi$}\label{Sec:v2flow_4pi}

Inspecting the time evolution of elliptic flow $v_2$ in $4\pi$ we notice, first of all, a very complex behavior, which depends in the form as well as in the magnitude on the energy. The contribution $\mathrm{d}v_2/\mathrm{d}t$ of collisions has a maximum around $t=\tmo$ whose amplitude increases strongly with energy
followed by a negative contribution for the two lower energies. The potential contribution shows two maxima and two minima. Shortly before the system reaches the highest density the contribution $\mathrm{d}v_2/\mathrm{d}t$ is negative (or at least close to zero) and shortly after it gets positive before it becomes again negative. Finally for times of $t=3-5\tmo$ we see a second maximum, which will be discussed later. 

The integrated flow reflects this complex time evolution. The contribution of the collisions to the integrated $v_2$ is negative at the lowest energy, slightly negative at the intermediate energy and positive at the largest energy whereas the potential contribution shows the opposite trend. It is positive at the lowest energy, slightly positive at the intermediate energy and negative at the highest energy. The total integrated $v_2$ follows the intricate interplay of the collisions and the potential. At the end of the evolution (here at 50 fm/c) it is close to zero at the lowest energy, at the intermediate energy it reaches 0.01 and at the highest energy it reaches 0.02.

\subsubsection{Elliptic flow $v_2$ at midrapidity}\label{Sec:flow_ymid}
So far we have discussed the change of $v_2$ in $4\pi$ acceptance, however, experiments mostly measure the midrapidity region, in which the evolution might be different. The Fig. \ref{fig:v2_ymid_time} thus shows the results at midrapidity defined by $|y|\leq 0.25$, independent of collision energy. 
The elliptic flow of the baryons at midrapidity is even more complex. The $\mathrm{d}v_2/\mathrm{d}t$ generated by the EoS again shows two maxima and two minima, while the differential change of $v_2$ of collisions has one pronounced minimum.
We further see that until $t=1.2 t_\mathrm{overlap}$ the positive contribution of the potential and the negative contribution of the collisions to $v_2$ almost cancel, therefore no net $v_2$ develops during the early time of the reaction despite of the fact that both amplitudes are strong. 
When the $v_2$ starts to develop at the time of maximal compression, both, collisions and potential, cause a negative $v_2$. Their relative contribution to the final $v_2$ is strongly energy dependent. At the lowest energy
the final $v_2$ is almost completely due to collisions, whereas at the highest energy the potential contributes more than half to the final $v_2$, which is nevertheless only half as large as at the lowest energy. This corresponds to the experimental observations shown in Fig. \ref{fig:experimental_flow}.




\subsubsection{Flow at kinetic freeze-out}\label{Sec:flow_frz}


The kinetic freeze-out is typically defined as the space-time point of a hadron's last interaction. In UrQMD the time of the last collisional interaction can be identified allowing to study the signed directed $v_1^*$ and elliptic $v_2$ flow at that time. After the last collision only potential interactions can modify the momentum of the nucleons. The $v_1$ and $v_2$ of the nucleons, which freeze out and which are are frozen out at t, is depicted in the bottom row of Figs. \ref{fig:v1_4pi_time} - \ref{fig:v2_ymid_time}. 

For both, the signed directed and elliptic flow, we observe that nucleons that freeze-out very early have a strongly negative flow value, having a magnitude much stronger than the final measured value. To understand this, we have to understand what determines the probability to freeze-out at any given location inside the fireball. Inside the fireball the local rate of freeze-out is given by the Pomeranchuk criterion relating by the ratio of the expansion rate $\Theta = \partial_\mu u^\mu$ with $u^\mu$ being the 4-velocity to the scattering rate of a given hadron species and momentum $\Gamma_i(p)$ \cite{Bondorf:1978kz,Hung:1997du,Inghirami:2021zja}. The escape probability is then proportional to the integrated optical depth along the trajectory of the particle, i.e. $P_\mathrm{esc} \propto \exp \left( -\chi \right)$ with $\chi = \int_t^\infty\mathrm{d} t^\prime \Gamma(x^\prime,t^\prime)$ \cite{Sinyukov:2002if,Knoll:2008sc}. Thus clearly, particles decoupling very early will only have a significant chance of decoupling if the optical depth along their respective trajectory is negligibly small. This is mostly satisfied if, by chance, the nucleons propagate towards lower density regions or straight into the vacuum. One further has to point out that, although the directed and elliptic flow values of nucleons decoupling very early are highly negative, not many nucleons at all decouple at this time. We therefore show additionally the integrated flow of nucleons that have frozen out until time $t$. 

One notices that at 50 fm/c the flow at freeze-out of all frozen out nucleons and the flow of the whole system (which includes the nucleons which already frozen out) a close but do not match exactly. The difference comes from the fact that nucleons which have decoupled kinetically, can further be de-/accelerated in the potential and thus change their flow or their rapidity at even later times. In addition, projectile and target spectator nucleons continue to interact by collisions. At midrapidity practically all nucleons are frozen out. 

At midrapidity the difference between the blue line in the bottom row and the black line in the middle row is therefore the potential contribution to $v_1^*$ and $v_2$ after the nucleons are frozen out.  It is remarkable that the integrated freeze-out $v_2$ at midrapidity at 50 fm/c is considerably lower (the values are -0.034, -0.046, -0.036 for the different energies) than the final $v_2$ of the system (the values are there -0.090, -0.083, -0.049). This means that 62\%, 44\% and 27\%, respectively,  of the finally observed  $v_2$ at midrapidity is due to the potential interaction after freeze-out. This change of $v_2$ has two origins: i) particles which have at freeze-out as well as in the final state a rapidity $|y|<0.25$ may change their $v_2$ due to the potential and ii) particles, which have at freeze-out $|y|<0.25$ but finally $|y|>0.25$, contribute to the final $v_2$ (the opposite can also be true but is rare). A closer inspection shows that the second process, caused
by the potential gradient acting on the nucleons in the bridge between projectile and target remnant, is the dominant one.

\begin{figure*} [t!hb]
    \centering
    \includegraphics[width=\columnwidth]{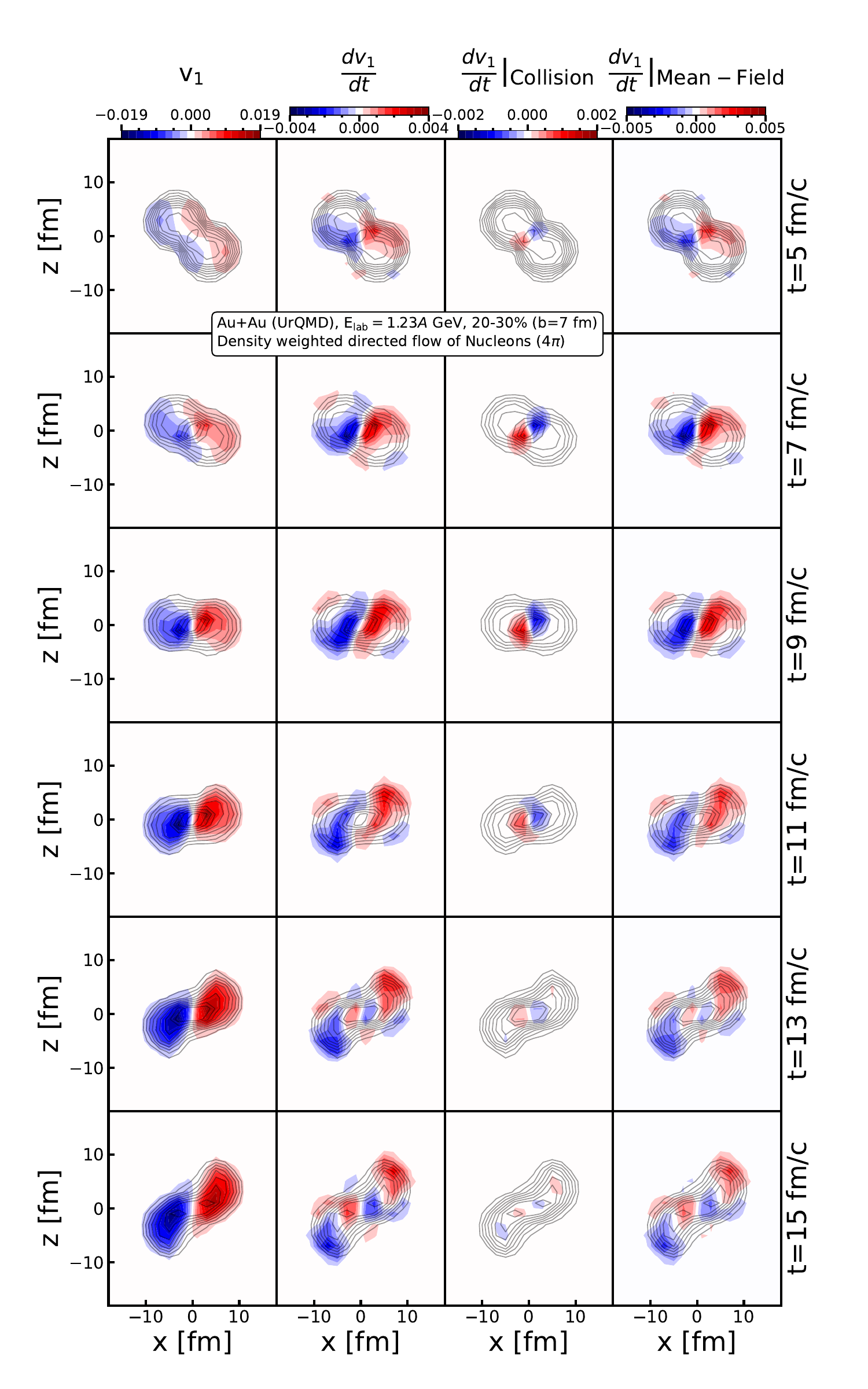}
    \includegraphics[width=\columnwidth]{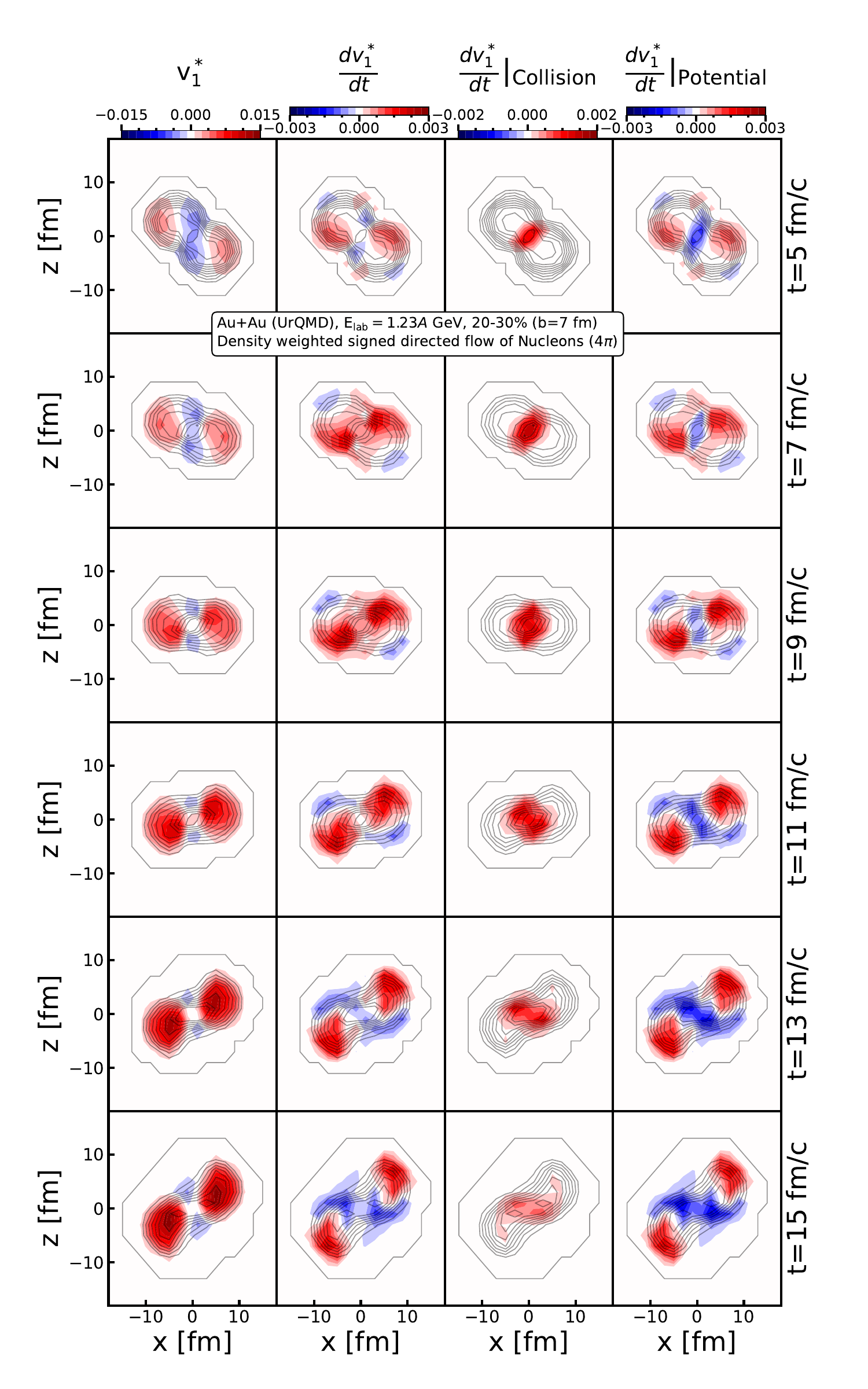}
    \caption{[Color online] The figure shows the time evolution of the density weighted directed flow $v_1$ (left panel) and of the density weighted signed directed flow $v_1^*$ (right panel) of baryons in the x-z plane in $4\pi$ at times t = 5, 7, 9, 11, 13 and 15 fm/c (from top to bottom, denoted on the right hand side) in 20-30\% (b=7 fm) peripheral Au+Au collisions at $1.23A$ GeV kinetic beam energy calculated with UrQMD with a hard Skyrme EoS. The columns show from left to right: the density weighted (signed) directed flow $v_1^{(*)}$, the change of (signed) directed flow with time $\mathrm{d}v_1^{(*)}/\mathrm{d}t$, the change due to collisions and the change due to the potential. The gray contour lines denote the density evolution of the system in the same phase space cut and plane. The magnitude of the flow and the change of the flow are denoted by the color bar shown at the top of each column. The color bars are symmetric and centered around 0.}
    \label{fig:v1_density_total}
\end{figure*}

\begin{figure*} [t!hb]
    \centering
    \includegraphics[width=\columnwidth]{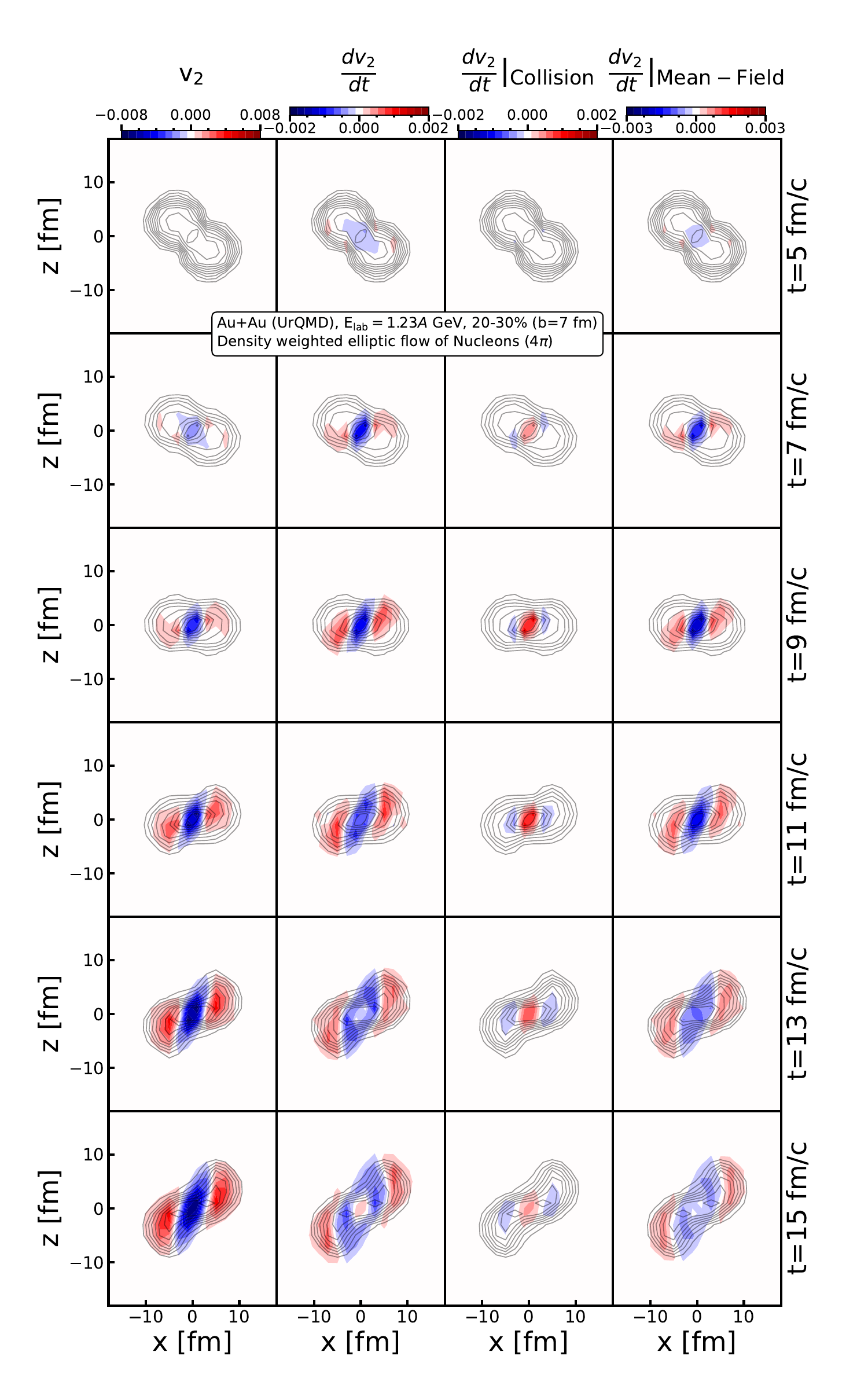}
    \includegraphics[width=\columnwidth]{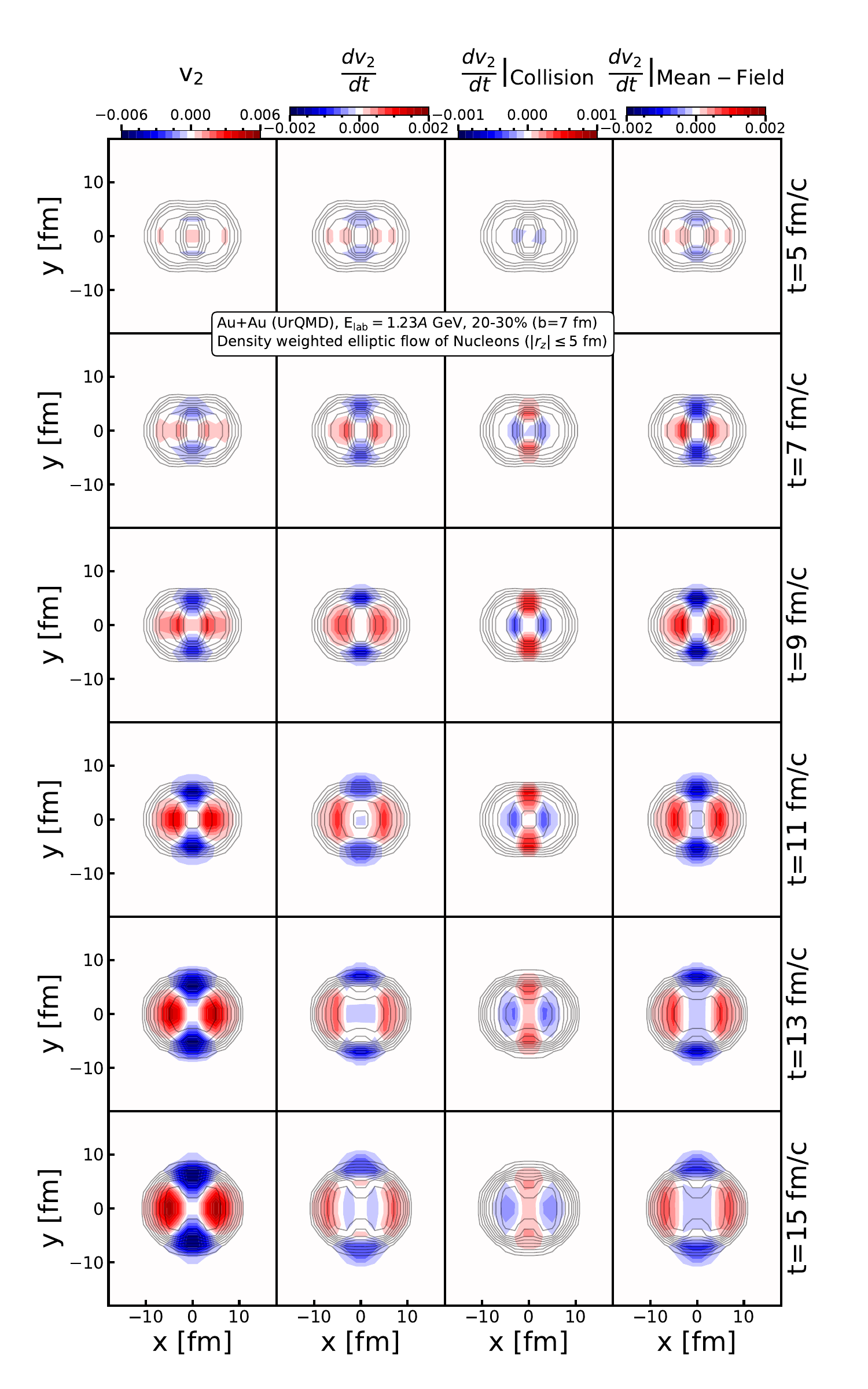}
    \caption{[Color online] The figure shows the time evolution of the density weighted elliptic flow $v_2$ of baryons in the x-z plane in $4\pi$ (left panel) and in the x-y plane at $|r_z|\leq5$ fm (right panel) at times t = 5, 7, 9, 11, 13 and 15 fm/c (from top to bottom, denoted on the right hand side) in 20-30\% (b=7 fm) peripheral Au+Au collisions at $1.23A$ GeV kinetic beam energy calculated with UrQMD with a hard Skyrme EoS. The columns show from left to right: the density weighted elliptic flow $v_2$, the change of elliptic flow with time $\mathrm{d}v_2/\mathrm{d}t$, the change due to collisions and the change due to the potential. The gray contour lines denote the density evolution of the system in the same phase space cut and plane. The magnitude of the flow and the change of the flow are denoted by the color bar shown at the top of each column. The color bars are symmetric and centered around 0.}
    \label{fig:v2_density_total_4pi}
\end{figure*}
\begin{figure*} [t!hb]
    \centering
    \includegraphics[width=\columnwidth]{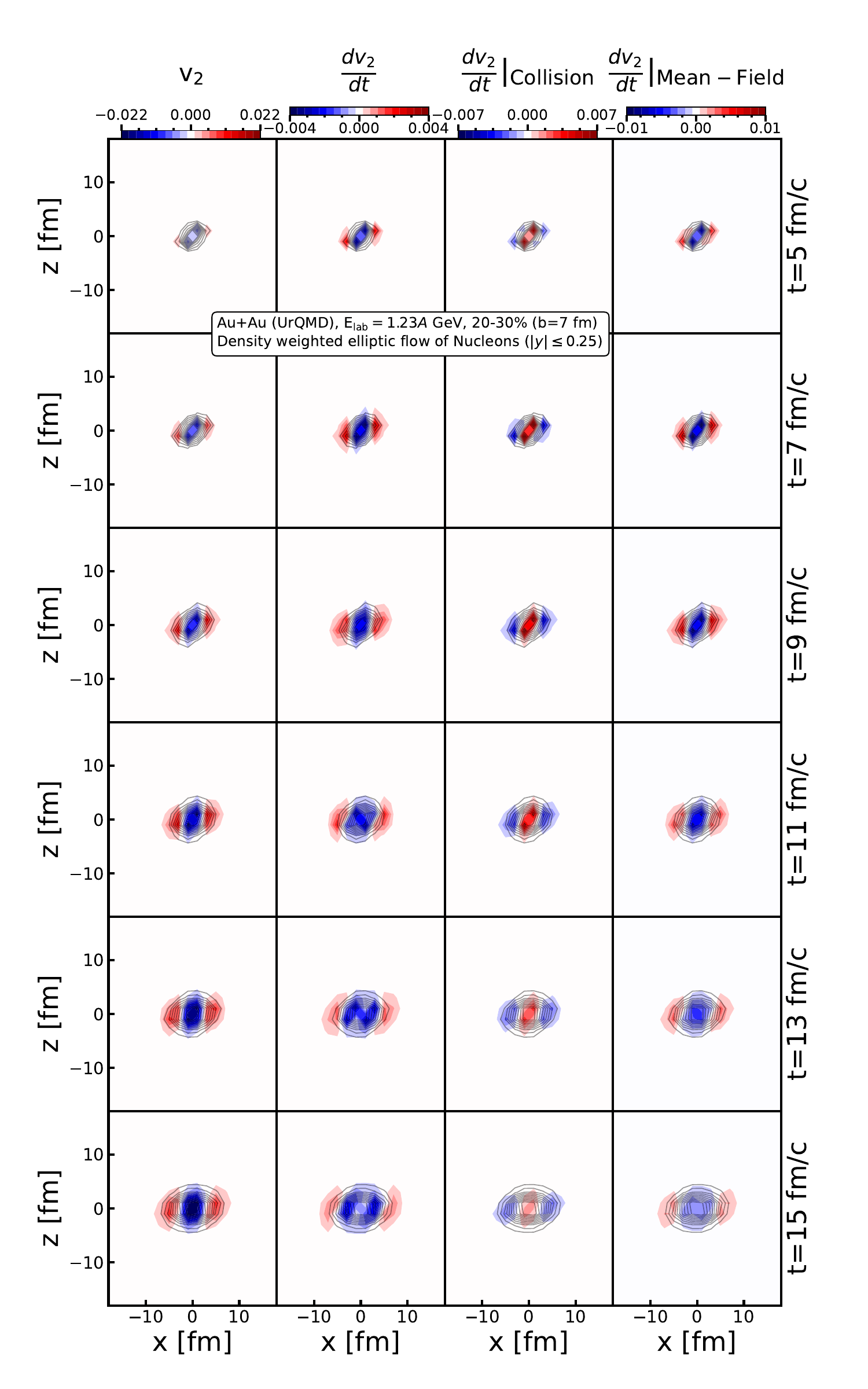}
    \includegraphics[width=\columnwidth]{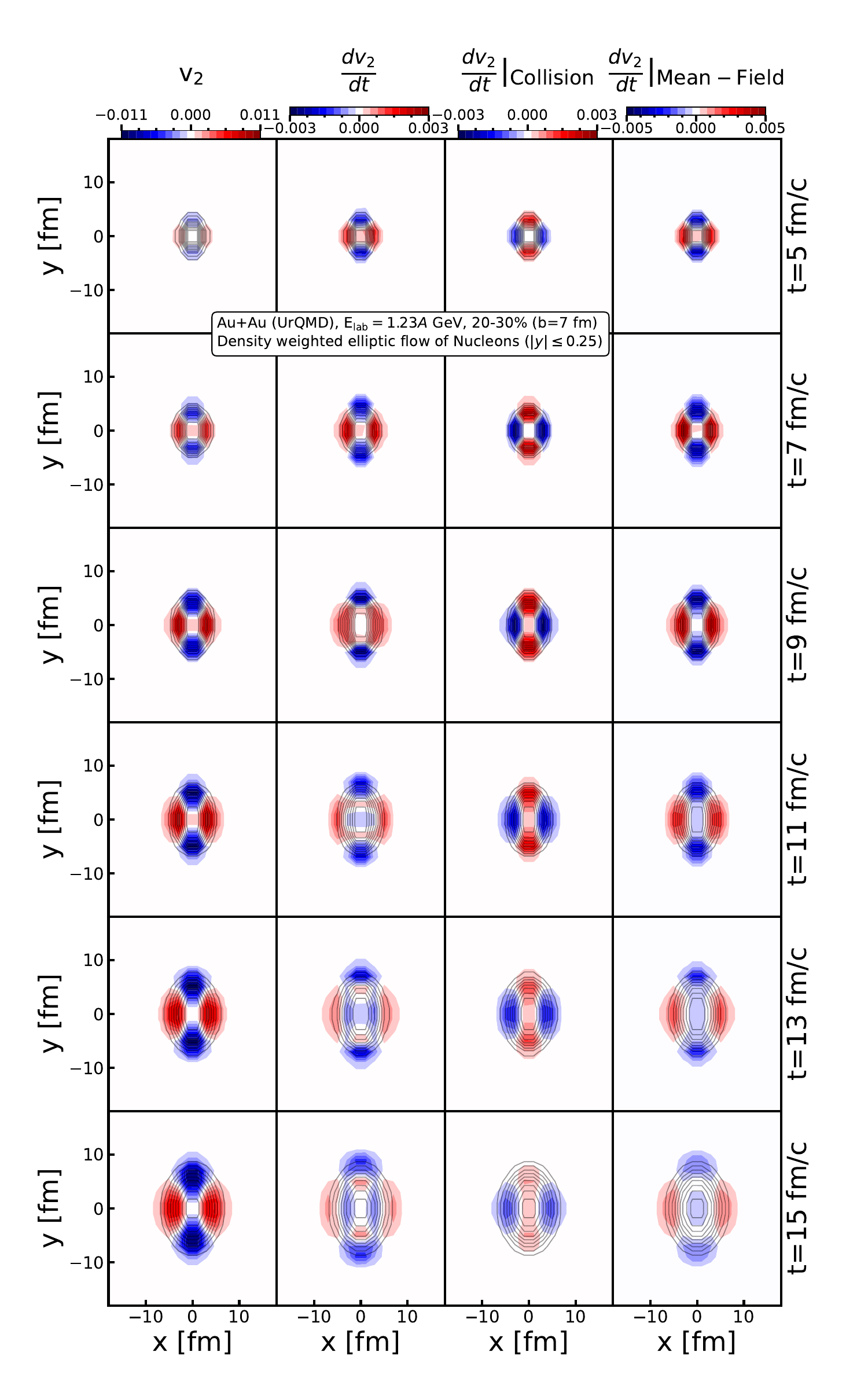}
    \caption{[Color online] The figure shows the time evolution of the density weighted elliptic flow $v_2$ of baryons in the x-z plane at $|y|\leq0.25$ (left panel) and in the x-y plane at $|y|\leq0.25$ fm (right panel) at times t = 5, 7, 9, 11, 13 and 15 fm/c (from top to bottom, denoted on the right hand side) in 20-30\% (b=7 fm) peripheral Au+Au collisions at $1.23A$ GeV kinetic beam energy calculated with UrQMD with a hard Skyrme EoS. The columns show from left to right: the density weighted elliptic flow $v_2$, the change of elliptic flow with time $\mathrm{d}v_2/\mathrm{d}t$, the change due to collisions and the change due to the potential. The gray contour lines denote the density evolution of the system in the same phase space cut and plane. The magnitude of the flow and the change of the flow are denoted by the color bar shown at the top of each column. The color bars are symmetric and centered around 0.}
    \label{fig:v2_density_total_ymid}
\end{figure*}

\subsection{Contributions of Potential and Collisions to the Flow}
The complex time evolution of the directed and the elliptic flow in the x-y and x-z plane has been shown in Figs. \ref{fig:v1_xz_xy} - \ref{fig:v2_xz_xy_weighted}. This qualitative presentation does, however, not explain the quantitative development of $v_1$ and $v_2$ as a function of time, displayed in Figs. \ref{fig:v1_4pi_time}, \ref{fig:v2_4pi_time} and \ref{fig:v2_ymid_time}. Further information is needed to understand the influence of collisions and potential on the experimentally measured flow coefficients and to reveal their physical origin.

To elucidate this further we present in Figs. \ref{fig:v1_density_total} - \ref{fig:v2_density_total_ymid} the time evolution of $v_n$ and $\mathrm{d}v_n/\mathrm{d}t$ for Au+Au reactions at E$_\mathrm{lab}=1.23A$ GeV. We display for different times (t = 5, 7, 9, 11, 13 and 15 fm/c) in the third column the change of the flow due to collisions, in the forth column that due to the potential and in the second column the sum of both, i.e. the total change of the flow. The first column displays the integrated value of the flow at this time point. In Fig. \ref{fig:v1_density_total} we display this for the directed flow $v_1$ (left panel) and the signed directed flow $v_1^*$ (right panel), both in the x-z plane, in Figs. \ref{fig:v2_density_total_4pi} and \ref{fig:v2_density_total_ymid}
for the elliptic flow in $4\pi$ and at midrapidity, respectively, both in the x-z (left panels) and x-y plane (right panels). The magnitude of the flow and the change of the flow are denoted by the color bar shown at the top of each column. The color bars are symmetric and centered around 0.

We start with the density weighted directed flow. We observe at early times ($t=0.5 \tmo$) that the high density, created at the overlap points of projectile and target, produces a directed flow in the range of the potential. We see as well that the collisions counteract immediately, trying to bring the system into a more equilibrated state. When time continues we see this counteraction of the collisions in the overlap zone increasing, it is absent, however, in the cold spectator matter where collisions are blocked by the Pauli principle. Finally, when the density in the bridge between projectile and target remnant decreases, the potential tries to increase the density in this bridge to normal nuclear matter density and accelerates by this the nucleons towards each other. This creates a negative $v_1$ increase of the nucleons located in vicinity of the bridge with a positive $r_x$ and a positive $v_1$ increase for those with a negative $r_x$. This late changes of $v_1$ are too weak to change the sign of the increase of $v_1$ but influence its final quantitative value. The change of $v_1$ at this late phase is dominated by the potential and is responsible for the second maximum in Fig. \ref{fig:v1_4pi_time}.

From the left panel of Fig. \ref{fig:v1_density_total} one can conclude that  $\mathrm{d}v_1/\mathrm{d}t$ due to collisions counteracts $\mathrm{d}v_1/\mathrm{d}t$ due to the potential, having the opposite sign. If one compares this result with the spatially integrated $v_1^*$, presented in Fig. \ref{fig:v1_4pi_time}, one has to take into account the difference between the directed flow $v_1$ and the signed directed flow $v_1^*$.  To allow for better comparison with the integrated value we thus also present the signed directed flow and its temporal variation in the x-z plane in the right panel of Fig. \ref{fig:v1_density_total}. We  observe that most of the negative parts of $v_1$ become positive if presented as the signed observable $v_1^*$, which is expected because these nucleons have negative rapidity. In the center of the system there remains, however, a region with a negative $\mathrm{d}v_1^*/\mathrm{d}t$, caused by the potential. Especially at the later times the potential is again seen to accelerate nucleons towards the lower density bridge forming in the center of the system after the two nuclei have passed each other. One very interesting feature that has not been present in the assessment of the directed flow without the sign of rapidity, is the appearance of anti-flow \cite{Snellings:1999bt,Brachmann:1999xt} denoted by the bluish region in the left most column. We observe this overall negative $v_1^*$ in the center of the participant zone up to the end of the simulations.
 
Fig. \ref{fig:v2_density_total_4pi} shows the contribution of the collision and the potential to the change of $v_2$ in the x-z plane in $4\pi$ (left figure) and in the x-y plane in $|r_z|\leq5$ fm (right figure). Even more than for $v_1$ we see here that collisions always counteract time-delayed the action of the potential. 
From the second and third column of the figure one observes that the $\mathrm{d}v_2/\mathrm{d}t$ of collisions and potential have locally an opposite sign and the magnitude of both is proportional to each other. This explains already a part of the observations in Figs. \ref{fig:v2_4pi_time}, \ref{fig:v2_ymid_time}. Around t=7 fm/c the potential gradient perpendicular to the reaction plane is roughly twice as large as in the reaction plane, where the spectators lower the gradient. This out-of-plane gradient squeezes participant nucleons into the y-direction, creating a negative $v_2$ \cite{LeFevre:2016vpp}. At the same time the gradients in the reaction plane accelerate spectator matter, which is in the range of the potential, provoking a positive $v_2$. As can be seen in Fig. \ref{fig:v2_4pi_time} the positive $v_2$ created by the potential in the reaction plane becomes around $t=t_\mathrm{overlap}$ even larger than the negative $v_2$ due to the squeeze. Due to the low excitation energy in the spectator matter collisions are Pauli blocked and much less effective to counterbalance the $v_2$ of the potential, therefore the space integrated collisional $\mathrm{d}v_2/\mathrm{d}t$ remains still positive. Later the density of the participant region  decreases leading to lower density gradients and a lower counterbalance by collisions there. The spectator matter is now more excited and therefore collisions (called shadowing)  take place more frequently, which counterbalance the $v_2$ created by the in-plane density gradient and yield a negative collisional $\mathrm{d}v_2/\mathrm{d}t$. Also the in-plane density gradient gets lower, leading to a fading away of the positive potential in-plane $\mathrm{d}v_2/\mathrm{d}t$  so that the overall $\mathrm{d}v_2/\mathrm{d}t$ due to the potential becomes negative again.

In the late stage of the reaction, when the matter bridge between projectile and target spectator starts to break, the attractive potential creates a negative $v_2$ by decelerating the participant matter in the reaction plane, as discussed for the time evolution of $v_1$. The density is then that low that collisions are not frequent and cannot counteract anymore. The overall final $v_2$ in $4\pi$ is positive, as we have seen in Fig. \ref{fig:v2_4pi_time}, but this positive $v_2$ is created at the very end of the reaction 
($t > 2 \tmo$) mostly by the nucleons from the dissolving matter bridge.

We proceed towards what is happening at midrapidity, shown in
Fig. \ref{fig:v2_density_total_ymid}. This is an interesting region because experimental data are available from low energy ($E_\mathrm{lab} \approx 100A$ MeV) up to the highest energies (Fig. \ref{fig:experimental_flow}).  
As already seen in the study of $v_1$ and $v_2$ in $4\pi$, also here collisions counteract the potential. The initial squeeze is, however, strongly reduced if one triggers on nucleons with $|y| \leq 0.25$ at that time because the majority of nucleons are not stopped when maximal compression is reached. Also Pauli blocking is less important because the phase space around midrapidity is little occupied. As can be seen from the left hand side of the figure, both, collisions and potential, strongly change $v_2$ but these changes are positive and negative in different space regions and compensate to a large extend. The net result is that at $t$=$\tmo$ the total $v_2$ is marginally positive, see Fig. \ref{fig:v2_ymid_time}. 

An important feature for the understanding of the time evolution of $v_2$ at midrapidity is the fact that the spectator matter hinders the spatial expansion of the midrapidity source in x-direction and therefore the transverse density profile gets elliptical with an increase of the ratio between the principal axis with time.
However, due to the presence of baryons with $|y|>0.25$, the in-plane density gradient is lower and therefore the negative increase of $v_2$ in out-of-plane direction outweighs the positive in-plane increase. Collisions counteract but more in-plane where the matter is located and contribute therefore globally as well a negative $v_2$. 

Later, when projectile and target separate, at around $2\tmo$, and when a matter bridge is built between them, we find still a quite high density of nucleons along this bridge but the density gradient and hence the force is not very large and both
contributions to the negative increase of $v_2$ fade away. Also in the out-of-plane direction the density and density gradient gets smaller and the  mean free path increases, leading to a small change of $\mathrm{d}v_2/\mathrm{d}t$.



These negative contributions at the late time of the reaction ($t> 1.5 \tmo$), are, as can be seen from Fig. \ref{fig:v2_ymid_time}, at the origin of the negative flow, which is observed in experiments. For the other energies, we observe qualitatively the same behavior at slightly shifted times. We can therefore conclude that the $v_2$, observed at midrapidity in experiment, is created late, starting around $t=1.5\tmo$. Before, the increase of $v_2$ is much stronger but has opposite sign in different space regions and the contributions from potential and collisions have locally as well an opposite sign because the collisions counteract the change of $v_2$ due the potential. 
The finally observed $v_2$ is hence a consequence of the fading away of the in-plane $v_2$ increase by the potential and of the expansion of the system in the out-of-plane direction, which makes collisions there less frequent. In addition, after their last collision particles are still de-/accelerated at the boundary between the matter bridge and the spectator remnant, decreasing the number of particles in the $|y|<0.25$ interval slightly. This mostly affects particles having a more positive $v_2$ than the average at midrapidity, hence decreasing the elliptic flow at $|y|<0.25$ in the final state below the value at kinetic freeze-out, i.e. their respective last collision.
As already mentioned, the finally observed $v_2$ is thus mostly due to the potential interaction. Squeeze-out and spectator absorption are both observed, especially in the early time of the heavy-ion interaction, but they are both only indirectly connected with the finally observed $v_2$ value.

\section{Summary of the generation of flow at SIS energies}
In this article we have addressed the long standing question whether the negative elliptic flow, measured at midrapidity in heavy-ion collisions at SIS18, RHIC-FXT and SIS100 energies, arises from shadowing or squeeze-out, the two mechanisms proposed so far. For this purpose we employed the Ultra-relativistic Quantum Molecular Dynamics (UrQMD) model in its current version with a hard Skyrme type potential. We have calculated the time evolution of 20-30\% (b=7 fm) peripheral Au+Au collisions at E$_\mathrm{lab}=0.6A$ GeV, E$_\mathrm{lab}=1.23A$ GeV and $\sqrt{s_\mathrm{NN}}=3.0$ GeV and scaled the time of the system by the respective time of full geometric overlap. The calculated flow and density evolution and the quantitative assessment of the variations of the directed flow and the elliptic flow due to potential and collisions, have revealed a highly intricate time evolution. 

For the directed flow $v_1$ we confirm that it is created by the density gradient (and hence by the potential) between the overlap zone (participants) and the spectators. The collisions want to bring the system back to a more equilibrated state.
Therefore locally the $v_1$, generated by the potential gradient, is partially counterbalanced by a $v_1$ created by collisions, which has the opposite sign.
The concrete flow pattern is influenced by the range of the potential. At the end of the reaction, when projectile and target separate again, the nucleons in the overlap zone produce a directed flow which is opposite to the flow of the spectators to which they belonged originally. This does not change the flow feature in a qualitative way but becomes important for a quantitative evaluation.

The elliptic flow $v_2$ is more complex than  previous publications suggest, which generally address the question whether it is due to a squeeze out in y-direction, what increases the average $\langle p_y \rangle$, or due to spectator shadowing what lowers $\langle p_x \rangle$. We find that in the early phase of the reaction (until $t=1.5\tmo$), when flow and squeeze should be active, no net elliptic flow is generated, because the elliptic flow, created by the potential, is counterbalanced by that created by collisions, in the direction of squeeze out (y-direction) as well as in the x-direction, where spectator absorption is expected. Thus neither the squeeze nor the shadowing is directly at the origin of the finally observed $v_2$.  The finally observed $v_2$ is generated late, at $t > 1.5 \tmo$, in $4\pi$ as well as at midrapidity, and is determined by the geometry when projectile and target nuclei separate again, still connected by a matter bridge. Potential interaction at the boundary of this matter 
bridge de-/accelerates also nucleons into or out of midrapidity $|y|<0.25$ even after the last collision.

The observed $v_2$ is therefore to a large extend created by the potential interaction but not related to an initial squeeze. This finding is confirmed if one compares the $v_2$ of the nucleons at freeze-out with their final $v_2$.

This complexity of the $v_1$ and $v_2$ flow, seen in the simulations of heavy-ion reactions, is challenging for the transport approaches because the quantitative value depends on the range of the potential interaction, on the nuclear Equation-of-State and the (elastic and inelastic) cross sections of nucleons. High quality flow data for different systems and different energies are therefore a very valuable information to improve these approaches, which are forced to employ parametrizations due to the lack of theoretical results or of experimental input. 

Our findings are thus highly relevant for experiments at GSI, RHIC-FXT and the upcoming FAIR facility, but also for experiments at FRIB, and for future theoretical interpretation of experimental data. The article further underlines the timeliness of precision measurements for deeper understanding of the Equation-of-State at large baryon densities.

\begin{acknowledgements}
The authors thank Christoph Hartnack and Arnaud Le F\`evre for fruitful discussions about the generation of harmonic flow. 
We further thank Behruz Kardan, Jan Steinheimer and Marcus Bleicher for inspiring discussion.
T.R. thanks J.A. and the people at SUBATECH for their kind hospitality during this project.
T.R. acknowledges support via the Procope Mobility Grant provided by the Ambassade De France En Allemagne (French Embassy in Germany) with grant number 185-DGM-E0402-03-001.
T.R. furhter acknowledges support through the Main-Campus-Doctus fellowship provided by the Stiftung Polytechnische Gesellschaft (SPTG) Frankfurt am Main and moreover thanks the Samson AG for their support.
This article is part of a project that has received funding from the European Union’s Horizon 2020 research and innovation programme under grant agreement STRONG – 2020 - No 824093.  
Computational resources were provided by the Center for Scientific Computing (CSC) of the Goethe University and the ``Green Cube" at GSI, Darmstadt. 
\end{acknowledgements}


\end{document}